\documentclass[11pt,a4paper]{article}
\pdfoutput=1
\usepackage{jheppub}
\usepackage{amsmath}
\usepackage{amsfonts}
\usepackage{amssymb}
\usepackage[tight]{subfigure}
\usepackage{hyperref}
\usepackage{cleveref}
\usepackage{multirow}
\newcommand{\beq}{\begin{equation}}
\newcommand{\eeq}{\end{equation}}
\newcommand{\beqa}{\begin{eqnarray}}
\newcommand{\eeqa}{\end{eqnarray}}
\newcommand{\bit}{\begin{itemize}}
\newcommand{\eit}{\end{itemize}}
\newcommand{\nx}{{N}_X}
\newcommand{\nxb}{\overline{N}_{{X}}}
\newcommand{\nm}{{N}}

\newcommand{\irrsixt}{\mathbf{16}}
\newcommand{\irrten}{\mathbf{10}}

\newcommand{\mc}[1]{\ensuremath{\mathcal{ #1 }}}
\title{Neutrino mass from M Theory SO(10)}
\author[a,b]{Bobby S. Acharya}
\author[a]{Krzysztof Bo\.{z}ek}
\author[c]{Miguel Crispim Rom\~{a}o}
\author[c]{Stephen F. King}
\author[a]{Chakrit Pongkitivanichkul}

\affiliation[a]{King's College,\\
WC2R 2LS, London, United Kingdom}
\affiliation[b]{International Centre for Theoretical Physics,\\
	I-­34151 Trieste, ITALY}
\affiliation[c]{University of Southampton,\\
SO17 1BJ, Southampton, United Kingdom}

\emailAdd{bobby.acharya@kcl.ac.uk}
\emailAdd{krzystof.bozek@kcl.ac.uk}
\emailAdd{m.crispim-romao@soton.ac.uk}
\emailAdd{s.f.king@soton.ac.uk}
\emailAdd{chakrit.pongkitivanichkul@kcl.ac.uk}
\abstract{We study the origin of neutrino mass from $SO(10)$ arising from
$M$ Theory compactified on a $G_2$-manifold. This is linked to 
the problem of the breaking of the extra $U(1)$ gauge group, 
in the $SU(5)\times U(1)$ subgroup of $SO(10)$,
which we show can achieved via a 
(generalised) Kolda-Martin mechanism. 
The resulting neutrino masses arise from a combination of the seesaw mechanism
and induced R-parity breaking contributions. The rather complicated neutrino mass
matrix is analysed for one neutrino family and it is shown how
phenomenologically acceptable neutrino masses can emerge.}
\keywords{$M$ Theory, $SO(10)$, Neutrino masses, Symmetry Breaking}

\begin{document}

\maketitle
\tableofcontents

%%%%%%%%%%%%%%%%%%%%%%%%%%%%%%%%%%%%%%%%%%%
%%%%%%%%%%%%%%%%%%%%%%%%%%%%%%%%%%%%%%%%%%%
%%%%%%%%%%%%%%%%%%%%%%%%%%%%%%%%%%%%%%%%%%%
%================================================================%
\section{Introduction and Motivation}

The discovery of neutrino mass and lepton mixing provides key evidence for 
new physics beyond the Standard Model (SM) \cite{King:2015aea,King:2003jb,Altarelli:2010gt,King:2013eh,King:2014nza}.
The seesaw mechanism \cite{Minkowski:1977sc,GellMann:1980vs,Yanagida:1979as,Mohapatra:1979ia,Schechter:1980gr} is an attractive possibilty to account 
for the origin of neutrino mass and lepton mixing in terms of right-handed neutrinos with large Majorana masses.
$SO(10)$ Grand Unified Theories (GUTs) \cite{Fritzsch:1974nn} predict such 
right-handed neutrinos which appear along with SM matter fields in a single $\mathbf{16}$ multiplet. When the $SO(10)$ gauge group is broken to that of the SM, neutrino mass is an inevitable 
consequence. In order to satisfy the constraint of gauge coupling unification, we shall here
assume low energy supersymmetry (SUSY) \cite{Martin:1997ns}. However to also account for gravity, one needs to 
go beyond gauge theories, and here we shall focus on an $M$ theory version of string theory \cite{Witten:1995ex,Horava:1995qa}.

Recently we showed how $SO(10)$ SUSY GUTs could emerge from 
$M$ Theory compactified on a $G_2$-manifold \cite{Acharya:2015oea}.
In this framework, discrete symmetry and Wilson lines \cite{Witten:2001bf} were used to prevent proton decay while maintaining gauge unification. In contrast to the $SU(5)$ version \cite{Acharya:2008zi,Acharya:2011te}, the Wilson line symmetry breaking mechanism 
in $SO(10)$ requires additional matter at the TeV scale, with the quantum numbers of an extra 
${\bf 16}_X$ plus  $\overline{\bf 16}_X$  \cite{Acharya:2015oea}. In addition,
there were a number of unresolved issues in this approach, notably the mechanism for breaking the extra gauged $U(1)_X$ which accompanies the SM gauge group after the Wilson line symmetry breaking mechanism in $SO(10)$. This gauge group is the usual one in the maximal $SO(10)$ subgroup $SU(5)\times U(1)_X$ \footnote{The $U(1)_X$ is also commonly called $U(1)_{\chi}$ in the literature.},
where $SU(5)$ embeds the SM gauge group.
The key point is that, since Abelian Wilson line symmetry breaking preserves the rank of the gauge group,
the $U(1)_X$ gauge group needs to be broken by some other mechanism in the low energy effective field theory. Since right-handed
Majorana neutrino masses can only arise once the $U(1)_X$ is broken, the origin of neutrino mass
is therefore linked to this symmetry breaking.

In this paper we address the problem of $U(1)_X$ breaking and neutrino masses arising from
the $SO(10)$ $M$ theory, following the construction in \cite{Acharya:2015oea}, although our approach to solving these problems may be more general than the specific example studied.
To break the $U(1)_X$ gauge symmetry, we employ a (generalised) Kolda-Martin mechanism 
\cite{Kolda:1995iw},
where higher order operators can break the symmetry, inducing vacuum expectation values (VEVs)
in the scalar right-handed neutrino components of both the 
matter ${\bf 16}$ and the extra ${\bf 16}_X$, as well as their
conjugate partners. The subsequent induced R-parity violation \cite{Barbier:2004ez} provides additional sources
of neutrino mass, in addition to that arising from the seesaw mechanism
\cite{Minkowski:1977sc,GellMann:1980vs,Yanagida:1979as,Mohapatra:1979ia,Schechter:1980gr}. The resulting $11\times 11$ neutrino mass
matrix is analysed for one neutrino family (nominally the third family) and it is shown how a
phenomenologically acceptable neutrino mass can emerge. We defer any discussion of flavour mixing 
to a possible future study of flavour from $M$ theory. Here we only show that symmetry breaking and
viable neutrino masses can arise within the framework of $M$ theory $SO(10)$, which is a highly non-trivial
result, given the constrained nature of $M$ theory constructions.

It is worth remarking that there are other alternative ways that have been proposed to study neutrino masses
in string theory, which are complementary to the approach followed here.
For example, it is possible to obtain large Majorana mass terms from instanton effects \cite{Acharya:2006ia,Blumenhagen:2006xt,Ibanez:2006da,Cvetic:2007ku,Buchmuller:2007zd}, large volume compactification \cite{Conlon:2007zza}, or orbifold compactfications of the heterotic string \cite{Buchmuller:2007zd}. However the origin of Majorana mass terms in $SO(10)$ has been non-trivial to realise from the string theory point of view. In GUTs all matter fields are unified in $\irrsixt$ multiplets whereas Higgs fields and triplet scalars are unified in $\irrten$. Since string theory does not predict 
light particles in 
representations larger than the adjoint, the traditional renormalisable terms involving $\mathbf{126}, \overline{\mathbf{126}}, \mathbf{210}$, e.g., $W \sim \mathbf{126}\:\irrsixt\:\irrsixt$, are not possible. The dominant higher order operators are quartic ones such as 
$W = \overline{\irrsixt}\:\overline{\irrsixt}\:\irrsixt\:\irrsixt$. 
Assuming that the supersymmetric partner of the right handed neutrino singlet gets a VEV, the Majorana mass is given by $M \sim \frac{\langle \widetilde{N} \rangle^2}{M_{PL}}$. However, the required values of neutrino mass imply $M > 10^{14}$ GeV, which gives $\langle \widetilde{N}\rangle \sim \sqrt{M m_{Pl}} \sim 10^{16}$ GeV.
The implementation of the seesaw mechanism \cite{Minkowski:1977sc,GellMann:1980vs,Yanagida:1979as,Mohapatra:1979ia,Schechter:1980gr} in other corners of string compactification
has also been discussed \cite{Faraggi:1990it,Faraggi:1993zh,Coriano:2003ui,Ghilencea:2002da}.

The layout of the remainder of the paper is as follows.
In section \ref{sec:review}, we will review the $SO(10)$ construction from $M$ Theory on $G_2$-manifolds, expanding the discussion in  \cite{Acharya:2015oea}. In section \ref{sec:sym_breaking_section}, the mechanism for $U(1)_X$ breaking will be given. The neutrino mass matrix will be analysed in section \ref{neutrino_section}, and the numerical results presented in section \ref{sec:numerical}. Finally we conclude in section \ref{sec:conclusion}.

%================================================================%
\section{SO(10) SUSY GUTS from $M$ Theory on $G_2$-manifolds} \label{sec:review}

%================================================================%

$M$ Theory compactified on a $G_2$-manifold leads to a 4 dimensional theory with $\mathcal{N}=1$ SUSY, where gauge fields and chiral fermions are supported by different types of singularities in the compactified space \cite{Acharya:2001gy,Acharya:2004qe}. Yang-Mills fields are supported on three dimensional subspaces of the extra dimensions, along which there is an orbifold singularity, while chiral fermions will be further localised on conical singularities localised on these three dimensional spaces and interact with the gauge fields.

One of the key features of $M$ Theory compactified on $G_2$-manifolds without fluxes is that it provides a  framework for generating hierarchies of mass scales. To understand the reason behind this notice that in $M$ Theory, the moduli fields, $s_i$, are paired with the axions, $a_i$, in order to form a complex scalar component of a superfield $\Phi_i$
\beq
\Phi_i = s_i + i a_i + \mbox{fermionic terms} \ .
\eeq
In the absence of fluxes, the axions enjoy an approximate shift-symmetry, which is remnant of the higher dimensional gauge symmetry, $a_i \to a_i + c_i $
where $c_i$ is an arbitrary constant. This Peccei-Quinn symmetry, in conjunction with holomorphicity of the superpotential, severely constrains the superpotential for the moduli. As such, terms which are polynomial in the moduli and matter fields are forbidden at tree-level in superpotential, appearing only in the K\"ahler potential.

In general 
non-perturbative effects such as instantons
break the above shift symmetry, and generate a non-perturbative
superpotential involving moduli and matter. Interactions will be generated by membrane instantons, whose actions are given by exponentials of the moduli. As the moduli stabilise and acquire VEVs, these exponentials will turn out to be small, and the VEV of the hidden sector superpotential naturally leading to a  generation of hierarchical masses at the GUT scale \cite{Acharya:2007rc}.
These ideas were used to construct the $G_2$-MSSM \cite{Acharya:2008zi,Acharya:2011te}, an $SU(5)$ SUSY GUT from $M$ Theory on a $G_2$ manifold with the MSSM spectrum. Here, we discuss an extention of the program to the $SO(10)$ GUT group \cite{Acharya:2015oea}, while referring to previous work on $G_2$ compactifications and consequent predictions for the parameters \cite{Acharya:2008hi,Acharya:2012tw}.

%================================================================%

%\subsection{SO(10) models\label{sec:so10models}}

In the remainder of this section,
we focus on the $SO(10)$ SUSY GUT from $M$ Theory on $G_2$ manifolds which we proposed in \cite{Acharya:2015oea}. %It was shown that Wilson line and discrete symmetry can be used to solve proton decay and mu problem in SU(5) GUT group models derived from M theory.
The breaking patterns of an abelian Wilson line are the same as the ones of an adjoint Higgs. The simplest case of a surviving group that is the most resembling to the SM is
\begin{equation}
SO(10) \to SU(3)_c \times SU(2)_L \times U(1)_Y \times U(1)_X  \ ,
\end{equation}
under which the branching rules of the GUT irreps read
\begin{align}
{\bf 10} : \ &H_u=({\bf 1},{\bf 2})_{\left(\frac{1}{2}, 2\right)} \oplus H_d=({\bf 1},{\bf 2})_{\left(-\frac{1}{2},-2\right)} \oplus D=({\bf 3},{\bf 1})_{\left(-\frac{1}{3},2\right)}\oplus \overline{D}=({\bf \overline{3}},{\bf 1})_{\left(\frac{1}{3},-2\right)}\ , \\
{\bf 16} : \ &L=( {\bf 1},{\bf 2})_{\left(-\frac{1}{2},3\right)}\oplus e^c=({\bf 1},{\bf 1})_{(1,-1)} \oplus  N=({\bf 1},{\bf 1})_{(0,-5)}\oplus u^c = ({\bf \overline{3}},{\bf 1})_{\left(-\frac{2}{3},-1\right)}\oplus \nonumber \\
&\oplus d^c =({\bf \overline{3}},{\bf 1})_{\left(\frac{1}{3},3\right)}\oplus Q= ({\bf 3},{\bf 2})_{\left(\frac{1}{6},-1\right)} \ ,
\end{align}
and the subscripts are the charges under $U(1)_Y \times U(1)_X$, which are normalised as
$Q_Y = \sqrt{\frac{5}{3}}Q_1 , \ Q_X = \sqrt{40}\tilde Q_{X} $,
where $Q_1$, $\tilde Q_X$ are $SO(10)$ generators.

The Wilson line can be conveniently represented as 
%in \eqref{eq:generalwilsonline},
\begin{equation}
\mathcal W = \exp\left[\frac{i 2 \pi}{N}\left(a Q_Y + b Q_X\right)\right]=\sum^\infty_{m=0} \frac{1}{m!} \left(\frac{i 2\pi}{N}\right)^m\left(a Q_Y + b Q_X\right)^m \ ,
\end{equation}
where the coefficients $a$, $b$ are constrained by the requirement that $\mathcal{W}^N = 1$ and specify the parametrisation of the Wilson line. Under the linear transformation
\begin{align}
\frac{1}{2}a+2 b \to \alpha \ , \\
\frac{1}{3}a-2b \to \beta\ ,
\end{align}
its action on the fundamental irrep then reads
\begin{equation}\label{eq:W10}
\mathcal{W} 10 = \eta^\alpha H_u \oplus \eta^{-\alpha} H_d \oplus \eta^{-\beta} D \oplus \eta^{\beta} \overline{D} \ ,
\end{equation}
where $\eta$ is the $N$th root of unity.

Likewise the Wilson line matrix acts on the 16 irrep as
\begin{equation}\label{eq:W16}
\mathcal{W}16=\eta^{-\frac{3}{2}\beta} L \oplus \eta^{ \alpha +\frac{3}{2}\beta} e^c \oplus  \eta^{-\alpha +
 \frac{3}{2}\beta} N \oplus  \eta^{-\alpha - \frac{1}{2}\beta} u^c \oplus  \eta^{ \alpha - \frac{1}{2}\beta} d^c \oplus  \eta^{\frac{1}{2}\beta} Q \ ,
\end{equation}
which could be simplified a bit further by replacing $\beta\to 2\beta$ without loss of generality, in order for the parameters to read as integers.

The effective discrete charges -- of different states on a chiral supermultiplet that absorbs Wilson line phases -- will be the overall charge of the discrete symmetry (common to all states belonging to the same GUT irrep) in addition to the Wilson line phases (different for each state inside the GUT irrep).

Having all the ingredients required to employ Witten's discrete symmetry proposal, we would like to have a consistent implementation of a well-motivated doublet-triplet splitting mechanism as it was done for $SU(5)$. Unfortunately the customary approach to the problem does not seem to work with $SO(10)$, as shown in \cite{Acharya:2015oea}. To understand this first notice that Witten's splitting mechanism can only work in order to split couplings between distinct GUT irreps. This is understood as $\mathcal W$ has the form of a gauge transformation of the surviving group and so it will never be able to split self bilinear couplings of a GUT irrep. For example, if one takes a $\textbf{10}$ with Wilson line phases to contain the MSSM Higgses, we can see from \cref{eq:W10} that both mass terms for the Higgses and coloured triplets are trivially allowed. We could consider that in order to split the Higgses, $H_u$ and $H_d$, from the coloured triplets -- $D$, $\overline D$ -- we would need to add another ${\bf 10}$, but it was shown that this cannot be achieved and so we are ultimately left with light coloured triplets.

In order to allow for light $D$, $\overline{D}$ we need to guarantee that they are sufficiently decoupled from matter to prevent proton-decay. To accomplish this, we can use the discrete symmetry to forbid certain couplings,
namely to {\it decouple $D$ and $\overline{D}$ from matter}.
Such couplings arise from the $SO(10)$ invariant operator $\mathbf{10}\ \mathbf{16}\ \mathbf{16}$, with $\mathbf{16}$ denoting the three $SO(10)$ multiplets, each containing a SM family plus right handed neutrino $N$. If $\mathbf{16}$ transforms as $\eta^{\kappa} \mathbf{16}$, the couplings and charge constraints are
\begin{align}
H_u \mathbf{16} \mathbf{16}\; : & \;2\kappa + \alpha + \omega = 0 \;\mbox{mod}\; N \\
H_d \mathbf{16} \mathbf{16}\; : & \;2\kappa - \alpha + \omega = 0 \;\mbox{mod}\; N \\
D \mathbf{16} \mathbf{16}\; : & \;2\kappa - \beta + \omega \neq 0 \;\mbox{mod}\; N \\
\overline{D} \mathbf{16} \mathbf{16} \; : & \;2\kappa + \beta + \omega \neq 0 \;\mbox{mod}\; N,
\end{align}
where we allow for up-type quark Yukawa couplings together with couplings to the right-handed neutrinos,
\beq
y_u^{ij}H_u^w \mathbf{16}_i \mathbf{16}_j \equiv y_u^{ij}H_u^w (Q_iu_j^c+L_iN_j +i\leftrightarrow j),
\label{yu}
\eeq
and similarly for down-type quarks and charged leptons.
%================================================================%
%\subsection{Color triplet mediated proton decay}

The couplings forbidden at a renormalizable tree-level by the discrete symmetry are generically regenerated from K\"ahler interactions through the Giudice-Masiero mechanism \cite{Giudice:1988yz}. While this provides the Higgsinos a TeV scale $\mu$-term mass, it also originates effective trilinear couplings with an $\mathcal{O}(10^{-15})$ coefficient. As these are generic, we need to systematically study their physical implications at low energies, such as proton-decay, R-parity violation, and flavour mixing.
%Given that in $M$ Theory there can be no irreps larger than the adjoint, it's natural to assume matter fields will be either in the ${\bf 16}$ or in the ${\bf 10}$ irreps, hence the natural candidates for proton decay mediator are only $D$ and $\overline{D}$.
%The $SO(10)$ allowed couplings that can induce proton-decay operators are schematically
%\beq
%W \supset \mathbf{10}^w\mathbf{16}^m\mathbf{16}^m = \lambda_1 \overline{D} d^c u^c + \lambda_2 D e^c u^c + \lambda_3 D Q Q + \lambda_4 \overline{D} Q L + \lambda_5 D N d^c \label{eq:proton} \ ,
%\eeq
%where for illustrative purposes we are working with only a single family, ${\bf 16}^m$. These interactions are generically present in the K\"ahler potential even if otherwise forbidden in the superpotential by the discrete symmetry of the compactified space.
%Since the experimental bound for proton lifetime is extremely long, these terms must be forbidden by the discrete symmetry as $D$ will have non-trivial Wilson line phases. However, moduli fields in general generate them with small coefficients as it was pointed out in \cref{sec:effcouplings}.

For proton decay, effective superpotential will be generate by the following K\"ahler potential
\beq
K \supset \frac{s}{m_{Pl}^2} \overline{D} d^c u^c + \frac{s}{m_{Pl}^2} D e^c u^c + \frac{s}{m_{Pl}^2} D Q Q + \frac{s}{m_{Pl}^2} \overline{D} Q L + \frac{s}{m_{Pl}^2} D N d^c + \mbox{h.c.} \ ,
\eeq
where we assume $\mathcal{O}(1)$ coefficients. As the moduli acquire non-vanishing VEVs, these become
\beqa
W_{eff} & \supset \lambda D Q Q  + \lambda D e^c u^c + \lambda D N d^c\label{eq:proton} + \nonumber \\
& +\lambda \overline{D} d^c u^c+\lambda \overline{D} Q L,
\eeqa
where we considering all couplings to be similar and taking one family for illustrative purposes. Notice that contrary to $SU(5)$ case, there is no extra contribution from rotation of $L$ and $H_u$ as the bilinear term $\kappa L H_u$ is not allowed by gauge invariance.

We estimate the scalar triplet mediated proton decay rate to be
\beq
\Gamma_p \simeq \frac{\left| \lambda^2 \right|^2}{16 \pi^2}\frac{m_p^5}{m_D^4} \simeq \left( 10^{42} \;\mbox{yrs}\right)^{-1} ,
\eeq
where we took the mass of the colour triplets to be $m_D \simeq 10^3$ GeV.

%================================================================%
%\subsection{Triplet scalars/fermions lifetime}
Another limit for triplet scalar comes from the cosmological constraints on its decay. As we have seen from proton-decay operators, triplet scalars can decay into quarks. If they start to decay during the Big Bang Nucleosynthesis (BBN) then nucleons could be disassociated, spoiling the predictions for light element 
abundances. We can estimate another limit on the triplet scalar mass by calculating its lifetime as it decay through the processes $D \rightarrow e^c u^c, Q Q, Q L, d^c u^c$, and we get
\beq
\Gamma  \simeq  \lambda^2 m_D \simeq ( 0.1 \;\mbox{sec})^{-1} ,
\eeq
which is approximately consistent with BBN constraint. They will also give interesting collider signatures due to their long-lived nature.
%================================================================%
\subsection{The vector-like family splitting}

Because the presence of a light vector-like pair coloured triplets spoils unification, we need a workaround that will preserve unification while keeping the presented doublet-triplet problem solution. We achieve this by considering the presence of extra matter that would form a complete GUT irrep with the coloured triplets, and hence restore unification. Unification constraints requires heavy states with equivalent SM gauge numbers, say $d^c_X$ and $\overline{d^c}_X$, that have to be subtracted from the spectrum. This can be achieved by adding a vector-like family pair, ${\bf 16}_X \overline {\bf 16}_X$, and splitting its mass terms using Wilson line phases. 

Furthermore, as the Wilson line breaking pattern is rank-preserving,  we still need to break the extra abelian gauge factor $U(1)_X$. This can be achieved if a scalar component of the right-handed conjugated neutrino pair of an extra vector-like family ${\bf 16}_X$, $\overline{\bf{16}}_X$ acquires VEVs. On top of this, this VEV can generate a Majorana mass for the matter right-handed conjugated neutrinos, providing a crucial ingredient for a type I see-saw mechanism.

In order to preserve gauge coupling unification, we notice that the down-type quarks -- $d^c_X$, $\overline{d^c}_X$ -- have the same SM quantum numbers as the coloured triplet pair -- $D$, $\overline{D}$ -- coming from  the ${\bf 10}$. We take ${\bf 16}_X$ to be localised along a Wilson line, and find that it
transforms under the discrete symmetry as
\begin{equation}
{\bf 16}_X \to  \eta^x \left( \eta^{-3\gamma} L \oplus \eta^{ 3\gamma+\delta} e^c \oplus  \eta^{3 \gamma - \delta} N \oplus  \eta^{-\gamma-\delta} u^c \oplus  \  \eta^{-\gamma +\delta} d^c \oplus  \eta^{\gamma} Q \right) .
\end{equation}
On the other hand, we let ${\bf {\overline{16}}}_X$ transform without Wilson line phases, ${\bf {\overline{16}}}_X \to \eta^{\overline x}\,  {\bf {\overline{16}}}_X$, and the condition for the mass term that will split the vector-like family is
\begin{equation}
\overline{d^c}_X d^c_X : x - \gamma + \delta + \overline{x} = 0 \mod N ,\
\end{equation}
whilst forbidding all the other self couplings that would arise from ${\bf 16}_X {\bf {\overline{16}}}_X$.
The  $d^c_X$, $\overline{d^c}_X$ quarks will then be naturally endowed a GUT scale mass through membrane instantons, provided that the singularities supporting ${\bf 16}_X$, $\overline{\bf{16}}_X$ are close enough to each other in the compactified space. The remaining states of ${\bf 16}_X$, $\overline{\bf{16}}_X$ will have a $\mu$ term of order TeV through the Giudice-Masiero mechanism.
The coloured triplets -- $D$, $\overline{D}$ -- and the light components of ${\bf 16}_X$, $\overline{\bf{16}}_X$ will effectively account for a full vector-like family. The light spectrum is then the one of MSSM in addition to this vector-like family, which in turn preserves unification, with a larger unification coupling at the GUT scale.

%================================================================%

%================================================================%

\subsection{R-parity violation\label{sec:RPV}}

Despite the existence of an effective matter parity symmetry inside $SO(10)$, the presence of a vector-like family will lead to R-parity violating (RPV) interactions though the VEV of the $N_X$, $\overline{N}_X$ components in the presence of moduli generated interactions. Furthermore, as we will see in detail in Section \ref{sec:sym_breaking_section}, the scalar component of the matter conjugate right-handed neutrino, $N$, will also acquire a VEV. These VEVs break $SO(10)$ and will inevitably generate RPV.
These interactions will mediate proton-decay, enable the lightest supersymmetric particle (LSP) to decay, and generate extra contributions to neutrino masses. In our framework RPV is generic, not only arising from allowed superpotential terms but as well from K\"ahler interactions involving moduli fields.

The interactions that break R-parity can either be trilinear or bilinear (B-RPV), and have different origins in our framework. The first contribution we can find comes from the tree-level renormalizable superpotential allowed by the discrete symmetry. Since we will encounter $\langle N \rangle \neq 0$, this means that even in a minimal setup, there will be an R-RPV contribution from matter Dirac mass coupling
\beq
W \supset y_\nu N H_u L  \ ,
\eeq
reading
\beq\label{eq:BRPVfromDirac}
W \supset y_\nu \langle N \rangle H_u L \ .
\eeq

Next we turn our attention to the K\"ahler potential, where interactions otherwise forbidden by the discrete symmetry might arise if there is a modulus with required charge. In such case, there is another contribution arising from the non-vanishing VEVs of $N_X$, $N$ $\overline N_X$ in conjugation with moduli VEVs. To see this, notice that in the K\"ahler potential there are generically interactions of the form
\beq
K \supset \frac{1}{m_{Pl}}N H_u L+\frac{s}{m_{Pl}^2} \overline N_X H_u L+\frac{s}{m_{Pl}^2} \overline N_X^\dagger H_u L +\mbox{ h.c.}\ ,
\eeq
where while the first term exists in zeroth order in moduli (otherwise there would be no neutrino Dirac mass in the superpotential), the last two are otherwise forbidden by the discrete symmetry, and $s$ denotes a generic modulus for each coupling. These terms will generate contributions to B-RPV as $N_X$, $N$ $\overline N_X$, $s$ acquire VEVs.

There are two types of contribution arising from the terms above. The first is generates through the Giudice-Masiero mechanism. As the moduli acquire VEVs, new holomorphic couplings will appear in the superpotential
\beq
W_{eff, 1} = \frac{m_{3/2}}{m_{Pl}} \langle N \rangle H_u L +0.1 \frac{m_{3/2}}{m_{Pl}} \langle N_X \rangle H_u L + 0.1 \frac{m_{3/2}}{m_{Pl}} \langle \overline N_X^\dagger \rangle H_u L \ ,
\eeq
where $m_{3/2} \simeq \mathcal{O}(10^{4})$ GeV, and since $s/m_{Pl} \simeq 0.1$ in $M$ Theory.
%, and the extra U(1) breaking involves $\langle N_X \rangle \neq 0$, $\langle \overline N_X \rangle \neq 0$, generating a contribution to the B-RPV coupling.
Notice that in principle we would also have a term in the K\"ahler potential involving $N$, but this can be found to be subleading in comparison to the term arising from the Dirac mass Eq. \eqref{eq:BRPVfromDirac}.

The second contribution arises if the F-terms of the fields $N_X$, $N$, $\overline N_X$ are non-vanishing. In this case, we expect the appearance of the contributions
\beq
W_{eff, 2} = \frac{\langle F_N \rangle}{m_{Pl}} H_u L+0.1\frac{ \langle F_{N_X} \rangle}{m_{Pl}} H_u L+0.1 \frac{ \langle F_{\overline{N}^\dagger_X}\rangle}{m_{Pl}}  H_u L \ ,
\eeq
and its magnitude will depend on how much F-breaking provoked by our symmetry breaking mechanism. Here we are considering that the case where $\overline N_X^\dagger H_u L$ cannot exist in the K\"ahler potential in zeroth order in a modulus field.
%Otherwise, this contribution will be enhanced by one order of magnitude as $\langle s \rangle/m_{Pl} \simeq 0.1$.

Putting all together, the B-RPV interactions account to the B-RPV paramter

\beq
W \supset \kappa H_u L
\eeq
with 
\beq
\kappa = \left(y_\nu+\frac{m_{3/2}}{m_{Pl}}\right) \langle N \rangle + 0.1 \frac{m_{3/2}}{m_{Pl}} \langle N_X \rangle + 0.1 \frac{m_{3/2}}{m_{Pl}} \langle \overline N^\dagger_X \rangle +\frac{\langle F_N \rangle}{m_{Pl}}  +  0.1 \frac{\langle F_{N_X}\rangle}{m_{Pl}} +  0.1 \frac{ \langle F_{\overline{N}^\dagger_X}\rangle}{m_{Pl}} \ ,
\eeq
and the relative strength of each contribution is model detail dependent, namely on neutrino Yukawa textures, symmetry breaking details, and F-flatness deviation.

In a similar manner, trilinear RPV couplings will be generated when $N$, $N_X$, $\overline N_X$, $s$ acquire VEVs.
In order to systematically study this, we notice that the trilinear RPV couplings come from the term
\begin{equation}
\mathbf{16}\ \mathbf{16}\ \mathbf{16}\ \mathbf{16} , \
\mathbf{16}_X \mathbf{16}\ \mathbf{16}\ \mathbf{16} , \
\mathbf{\overline{16}}_X^\dagger \mathbf{16}\ \mathbf{16}\ \mathbf{16}
\end{equation}
as the scalar component of $N_X$, $N$ acquires non-vanishing VEVs. Notice that the last term lives in the K\"ahler potential.
These are made forbidden at tree-level using the discrete symmetry of the compactified space. However, just like the $\mu$ terms and the B-RPV terms shown above, these terms will in general be present in the K\"ahler potential and will effectively be generated as the moduli acquire VEVs.
This happens again through the Giudice-Masier mechanism and we will find
\beq
\mathcal{O} \left(\frac{m_{3/2}}{m_{Pl}^2} (\langle N \rangle + \langle N_X \rangle+\langle \overline{N}_X^\dagger \rangle)\right) \{L L e^c,\ LQ d^c,\ u^c d^c d^c\}  ,
\eeq
where $m_{3/2}/m_{Pl}\simeq \mathcal{O}(10^{-14})$. The apparent suppression of trilinear RPV is understood as these terms can only be generated by non-renormalizable terms in an $SO(10)$ context.

Similarly to the B-RPV case, there will be further contributions if the F-terms of $N_X$, $N$, $\overline N_X$ are non-vanishing. Namely we find
\beq
\mathcal{O}  \left(\frac{\langle F_N \rangle + \langle F_{N_X} \rangle + \langle F_{\overline{N}_X^\dagger}\rangle}{m_{Pl}^2}\right) \{L L e^c,\ LQ d^c,\ u^c d^c d^c\} ,
\eeq
and again we expect these to be sub-leading even if the F-terms are not vanishing.

We see then that the values of all RPV coupling are strictly related to the details of the breaking mechanism employed to break the extra $U(1)_X$. This will be studied in great detail in Section \ref{sec:sym_breaking_section}. Furthermore, the bilinear B-RPV term generates a contribution to the physical neutrino masses \cite{Banks:1995by,Barbier:2004ez}. The complete picture of neutrino masses, including B-RPV operators, will be discussed in Section \ref{neutrino_section}.

We can study now some direct effects of RPV in the dynamics of our class of models. Under the assumption that $\kappa \ll \mu$, performing a small rotation, of
$\mathcal{O}(\kappa/\mu)$, in $(H_d,L)$ space, the last term can be absorbed $\mu H_dH_u$.
As a consequence, the first two terms will be enhanced by the Yukawa couplings $y_eH_dLe^c$, etc., leading to
\beq
W \supset  y_e \frac{\kappa}{\mu}  L L e^c +  y_d \frac{\kappa}{\mu}  LQ d^c +  \lambda  \frac{v}{m_{Pl}} u^c d^c d^c ,  \label{eq:RPVrot}
\eeq
and we have dropped the $\mathcal{O}(1/m_{Pl})$ contributions to the first two terms since now the Yukawa rotated contributions are much larger.
Also, we kept the last term with the parametrization $v$ describing all contributions. These will be very small, for example in the case the VEVs are high-scale, $\langle N_X \rangle \simeq 10^{16}$ GeV, the trilinear RPV coupling strength is of $\mathcal{O}(10^{-16})$. A direct consequence of this result is that proton decay will be slow, even when the $\Delta L = 1$ terms are enhanced.

While the proton is relatively stable, the enhanced terms will provide a decay channel for the LSP, which is now unstable.
In the limit that we can take the final states to be massless, and considering that the LSP is a neutralino mainly composed of neutral gauginos, the LSP lifetime through the decay $\tilde \chi^0 \to d^cQL$ can be estimated from a tree-level diagram involving a virtual $\tilde{d}^c$ with mass $m_0$,\footnote{See, for example, the diagrams in \cite{Martin:1997ns}.} 
\begin{equation}
\tau_{LSP} \simeq \left(3.9 \times 10^{-15} \right)\left(\frac{\mu}{g_w y_d \kappa}\right)^2\left( \frac{m_0}{10 \ \mbox{TeV}}\right)^4 \left( \frac{100 \ \mbox{GeV}}{m_{LSP}}\right)^5 \sec,
\end{equation}
where $g_w$ is a weak gauge coupling. The LSP lifetime is bounded to be either $\tau_{LSP} \lesssim 1$ sec or $\tau_{LSP} \gtrsim 10^{25}$ sec \cite{Banks:1995by,Dreiner:1997uz}, from Big Bang Nucleosythesis (BBN) and indirect Dark Matter (DM) experiments, respectively. If we take $m_{LSP} \simeq 100\mbox{ GeV}$, $m_0 \simeq 10 \mbox{ TeV}$, $y_d=y_b \simeq 10^{-2}$, $g_w \simeq 0.1$, we find that the VEV $v_X$ is constrained to be either
\begin{align}
\kappa & \gtrsim 6 \times 10^{-2} \mbox{ GeV} \label{eq:kLSPBBN} \\
\ {\rm or} \  \kappa &\lesssim 2 \times 10^{-14}  \mbox{ GeV} ,
\end{align}
for a short- and long-lived LSP, respectively. In the above estimate we used the fact that the decay involving the bottom Yukawa is the largest contribution to the decay width.

We can use the above result to infer some parametric dependence on the scale of the $U(1)_X$ breaking. If we have the leading contribution to the B-RPV coupling to be $\kappa \simeq \langle N_X \rangle  \lambda \Rightarrow \langle N_X \rangle \gtrsim 10^{12}$ GeV. In this case, the LSP is too short lived to be a good DM candidate, but decays quickly enough to not spoil BBN predictions. On the other-hand, a low-scale VEV is bound to be $\langle N_X \rangle \lesssim 1$ GeV in order to allow for a long-lived LSP. This would imply the abelian gauge boson associated with extra $U(1)_X$ to be light, $m_{Z^\prime} < \mathcal{O}(1)$ GeV. This last scenario is completely excluded from experimental searches.

The lack of a good DM candidate in the visible sector indicates us that DM is realised elsewhere. For instance, it has been recently suggested that in the context of String/$M$ Theory, the generic occurrence of hidden sectors could account for the required DM mechanics \cite{Acharya:2016fge}.

%================================================================%

\subsection{The see-saw mechanism}

The relevance of the bounds on the rank-breaking VEV is only fully understood when studying the details of symmetry breaking mechanism and neutrino masses. For example, if we start with an $SO(10)$ invariant theory the Yukawas are unified for each family leading to at least one very heavy Dirac neutrino mass, $m_\nu^D$. However, if the right-handed conjugated neutrino has a heavy Majorana mass, then the physical left-handed neutrino mass will be small through a type I see-saw mechanism. In order to accomplish this, one has to allow the following terms in the superpotential
\beq
W \supset y_{\nu} H_u L N + M N N ,
\eeq
where $y_\nu$ are the neutrino Yukawas, $L$ the matter lepton doublets, $N$ the right-handed conjugated neutrino, and $M$ its  Majorana mass, which we take $M \gg m_\nu^D = y_\nu \langle H_u \rangle$. With the above ingredients, a mostly left-handed light neutrino will have a physical mass
\beq
m^\nu_{phy} \simeq - \frac{(m_\nu^D)^2}{M} .
\eeq
One of the most appealing features of $SO(10)$ models is that each family is in a $\bf{16}$ which includes a natural candidate for the right-handed conjugated neutrino, the $N$. In order to employ a type I see-saw mechanism, we need to generate a Majorana mass term for the matter right-handed conjugated neutrino through the operator $W\supset \overline{\bf 16}_X\overline{\bf 16}_X {\bf 16}\: {\bf 16}$ \footnote{Given that in $M$ Theory one does not account for irreps larger than the adjoint, this is the lowest order term that can generate a right-handed neutrino Majorana mass.} leading to the operator
\beq
\frac{1}{m_{Pl}} \overline{N}_X \overline{N}_X N N\ ,
\eeq
from which the Majorana mass for the (CP conjugated) right-handed neutrino field $N$
is emerges as
\beq
M \simeq \frac{\langle \overline{N}_X \rangle^2}{m_{Pl}} \ .
\eeq
We can now relate the bounds on the value of the D-flat VEVs
$\langle \overline{N}_X \rangle = \langle N_X \rangle$ from both RPV and the 
requirement of a realistic see-saw mechanism. Since the physical neutrino mass in type I see-saw mechanism is given by
\beq
m_\nu^{phy} \simeq \frac{(m_\nu ^D)^2}{M} \ ,
\eeq
assuming $m_\nu ^D \simeq \mathcal{O}(100\mbox{ GeV})$, and knowing that the upper bound on the neutrino masses $m_\nu^{phy} \lesssim 0.1$ eV, one finds
\beq
M \gtrsim 10^{14} \mbox{ GeV} \Rightarrow
\langle {N}_X \rangle \gtrsim  10^{16} \mbox{ GeV}.
\eeq
The above argument suggests that we need to break the $U(1)_X$ close to the GUT scale. Since the Wilson line breaking mechanism is rank-preserving, we need to look for an alternative solution. Although the neutral fermion mass matrix will be considerably more intricate, obscuring the relations and hierarchies amongst different contributions to the neutrino masses, the above estimate motivates
the need for a high-scale $U(1)_X$ breaking mechanism.

%================================================================%
\subsection{Effective light families}
For a simple SUSY $SO(10)$ model where each family is unified into a single irrep with universal soft masses, it is well known that
electroweak symmetry is difficult to break \cite{Hall:1993gn,Rattazzi:1994bm,Murayama:1995fn,Baer:1999mc,Auto:2003ys,Baer:2009ie}. Since the two Higgs soft masses are unified at GUT scale and have similar beta function due to Yukawa unification, either both masses are positive at electroweak scale and symmetry is not broken or both masses are negative and the potential becomes unbounded from below. Another aspect of Yukawa unification problem lies in the fact that low energy spectrum of quarks and leptons requires some degree of tuning in parameter space when their RG runnings are considered.

The EWSB and Yukawa textures issues are naturally solved if each family is not contained in one single complete ${\bf 16}$, but is instead formed of states from different Ultra Violet (UV) complete ${\bf 16}$s. In order to implement this in our framework, first we assume the existence of multiple ${\bf 16}$ with independent and different Wilson Line phases, alongside the existence of multiple ${\overline{\bf 16}}$. Second, we employ Witten's proposal to turn on some vector-like masses such that three effective light ${\bf 16}$ survive. Since in $M$ Theory the strength of the Yukawa couplings is given by membrane instantons, and are therefore related to distances between the singularities supporting the respective superfields, by constructing effective families from different UV ${\bf 16}$s one can obtain different Yukawa couplings within each family.

Such solution can be achieved if one considers $M$ complete ${\bf \overline{16}}_j$ and $M+3$ complete ${\bf 16}_i$ UV irreps. Allowing for masses between different states of these UV irreps to appear, one has schematically the mass terms in the superpotential
\begin{equation}
\mathbf{16}_i \mu_{ij} \overline{\mathbf{16}}_j,
\end{equation}
but since $i=1,...,M$ while $j=1,...,M+3$ the mass matrix $\mu_{ji}$ can only have at most rank $M$ and hence there will be three linear combinations composing three ${\bf 16}$ that will remain massless. If these masses are truly $SO(10)$ invariant, i.e.
\begin{equation}
\mathbf{16}_i \mu_{ij} \overline{\mathbf{16}}_j = \mu_{ij} \left( Q_i \overline Q_j + L_i \overline L_j + \ldots\right),
\end{equation}
each effective light family will be $SO(10)$ invariant. Consequently each family will retain unified Yukawa textures, and so this does not solve our problem of splitting the Yukawa couplings within each family.

However, Witten's proposal endows our framework with a GUT breaking discrete symmetry which can be employed to ensure that the superpotential mass matrices between the UV states
\begin{equation}
\mu^Q_{ij}  Q_i \overline Q_j +\mu^L_{ij}  L_i \overline L_j + \ldots,
\end{equation}
are not the same, leading to different diagonalisations of $Q$, $L$, etc which in turn break the Yukawa $SO(10)$ invariance. In order to accomplish that, take for example that the ${\bf 16}_i$ absorb distinct and independent Wilson line phases, while $\overline{\bf 16}_j$ do not, i.e. the UV irreps will transform under the discrete symmetry as
\begin{align}
{\bf 16}_i &\to\eta^{m_i}\left(\eta^{-3\gamma_i} L_i \oplus \eta^{ 3\gamma_i+\delta_i} e_i^c \oplus  \eta^{3 \gamma_i - \delta_i} N_i \oplus  \eta^{-\gamma_i-\delta_i} u_i^c \oplus  \eta^{-\gamma_i +\delta_i} d_i^c \oplus  \eta^{\gamma_i} Q_i\right) \label{eq:16icharges}\\
\overline{\bf 16}_j &\to \eta^{\overline m_i}\overline{\bf 16}_j \label{eq:16bjcharges},
\end{align}
and look for solutions for the discrete charges where different states have different mass matrices. Since explicit examples can only be given by solving extensive modular linear systems, which are computationally prohibitive, a fully working example with three light-families is not provided.

%================================================================%
\section{$U(1)_X$ Breaking scenarios and mechanisms} \label{sec:sym_breaking_section}

In this section we are interested in implementing a symmetry breaking mechanism for the extra $U(1)_X$ in which the breaking VEV is stabilised at high values, more or less close to the GUT scale. In order to do so, we will look into the D-flat direction of the potential that breaks the extra $U(1)_X$. It was shown \cite{Drees:1986vd,Kolda:1995iw} that in the D-flat direction, non-renormalisable operators can provide such scenario. In its simplest inception, the Kolda-Martin
mechanism \cite{Kolda:1995iw} relies on a vector-like pair which lowest order term allowed in the superpotential is non-renormalizable
\begin{equation}
 W = \frac{c}{m_{Pl}} (\Phi \bar \Phi)^2
\end{equation}
and alongside the soft-term Lagrangian
\begin{equation}
 -\mc{L}_{soft} = m_\Phi^2 |\Phi|^2 + m_{\bar\Phi}^2 |\bar{\Phi}|^2,
\end{equation}
it is immediate to find that along the D-flat direction the potential has a non-trivial minimum which fixes the VEVs at a high scale
\begin{equation}
\Phi^2 = \sqrt{-\frac{(m_\Phi^2+m_{\bar\Phi}^2)m_{Pl}^2}{12 c}} ,
\end{equation}
where if we take $m \simeq 10^4$ the VEVs are estimated at $\Phi \simeq 10^{11}\mbox{ GeV}$.

There are some caveats to this mechanism as presented above. First, there is significant F-breaking as $\langle F \rangle \simeq \mc{O}(10^{15})\mbox{ GeV}$. While this is not a problem if the vector-like family does not share gauge interactions with ordinary matter, in our case non-vanishing F-terms will originate undesirable interactions, c.f. Section \ref{sec:RPV}. We shall therefore focus on F-flat solutions.

Second, the mechanism is not complete in the absence of the full soft-terms Lagrangian, which has to include
\begin{equation}
-\mc{L}_{soft} \supset C \frac{1}{m_{Pl}} \Phi^2 \bar{\Phi} ^2 + \mbox{ h.c.} .
\end{equation}
As we estimate $C\simeq\mc{O}(m_{3/2})$ at the GUT scale from the SUGRA \cite{Brignole:1997dp}, at the VEV scale this term is competing with the non-renormalisable terms in the potential arising from the superpotential, and therefore cannot be ignored.

Finally the model presented differs from ours as $\mu$-terms are generically generated by moduli VEVs even if they are disallowed by the discrete symmetry of the compactified space.

In order to proceed, we turn to a more complete version of the mechanism. To do so, we include the $\mu$-term
\begin{equation}
W = \mu \Phi \bar{\Phi}+\frac{c}{m_{Pl}} (\Phi \bar \Phi)^2
\end{equation}
and the more complete soft Lagrangian,
\begin{equation}
 -\mc{L}_{soft} = m_\Phi^2 |\Phi|^2 + m_{\bar\Phi}^2 |\bar{\Phi}|^2 - (B\mu \Phi \bar \Phi + \mbox{ h.c.})+  \frac{C}{m_{Pl}} \Phi^2 \bar{\Phi} ^2 + \mbox{ h.c.}  .
\end{equation}
Due to the presence of the $\mu$-term, the F-term
\begin{equation}
F_{\Phi} = \mu \bar \Phi + \frac{2 c }{m_{Pl}} \Phi \bar \Phi ^2
\end{equation}
can be set to zero for two different field configurations
\begin{equation}
 F_\Phi =0 \Rightarrow \left\{
 \begin{array}{l}
 \bar \Phi  =0 \\
 \Phi \bar \Phi = - \frac{\mu m_{Pl}}{2 c}
 \end{array}
 \right.
\end{equation}
and the non-trivial VEV can be estimated. Taking $\mu \simeq \mc{O}(10^3)$ GeV, this leads to $| \Phi | = 10^{10.5}$ GeV. This looks very similar to the original Kolda-Martin case, with the exception being that the F-term can vanish, and the parametric dependence on the VEV is now on $\mu$ instead of a soft-mass. In general there might be a non-SUSY preserving vacuum elsewhere in field space, but we will work under the assumption that the SUSY vacua discovered with this approach are at least stable enough to host phenomenologically viable models.

We wish to assess if we can minimise the potential in this SUSY-preserving field configuration. For that, we need to check if the above field configuration will also extremise the soft-term Lagrangian. To see this we take
\begin{equation}
- \partial_\Phi \mc{L}_{soft} = m^2_\Phi \Phi^* - B\mu \bar \Phi + \frac{2 C}{m_{Pl}} \Phi \bar \Phi^2 =0
\end{equation}
and, in the limit the VEVs are real, we find a trivial and a non-trivial solutions
\begin{equation}
- \partial_\Phi \mc{L}_{soft}  = 0 \Rightarrow \left\{
\begin{array}{l}
\Phi  =0 \\
\Phi ^2 = - \frac{(m_\Phi^2-B\mu) m_{Pl}}{2 C}
\end{array}
\right.
\end{equation}
and the second one seems very similar to the non-trivial configuration derived through the F-term. In fact, both conditions can be met. To see this, we re-parametrise the soft-terms by factoring out their dimensionful dependence on $m_{3/2}$
\begin{align}
	B\mu &= m_{3/2} \mu b \\
	C & = m_{3/2} \tilde c \\
	m_\Phi & = m_{3/2} a ,
\end{align}
where $a$, $b$, $\tilde c$ are dimensionless, and from SUGRA formulae they are $\mc{O}(1)$ at the GUT scale. Of course they will evolve with the scale through RGE evolution, so they need not to be always of the same order. The condition that both the F-flatness and soft-term stabilisation are jointly achieved boils down to be a relation between parameters
\begin{equation}
\frac{\tilde c}{c} = \frac{2 a m_{\Phi} -\mu b}{\mu} ,
\end{equation}
which is generically valid.

In order for the above non-trivial VEV be a minimum, we need the trivial VEV solution to account for a maximum. This is to say that the mass matrix for the system $(\Phi, \bar \Phi^*)$ evaluated at the origin has a negative eigen-value. In our case this accounts for allowing its determinant to be negative
\begin{equation}
(|\mu|^2+m^2_\Phi)(|\mu|^2+m^2_{\bar \Phi})-B\mu^2<0.
\label{eq:CondTrivialMaximum}
\end{equation}

We notice as well that the above discussion can be immediately extended for the case that the lowest order non-renormalisable term allowed by the discrete symmetry
\begin{equation}
W \supset \frac{c}{m_{Pl}^{2n-3}}(\Phi \bar \Phi)^n \Rightarrow \Phi \simeq (\mu m_{Pl}^{2n-3})^{\frac{1}{2n-2}}
\end{equation}
happens for $n \geq 2$, and not only for $n=2$. Even so, the presented implementation of the Kolda-Martin mechanism only accounts for a vector-like pair of superfields, while in our case the system breaking the extra $U(1)_X$ is composed of $N$, $N_X$, $\overline N_X$ states.

Therefore, we want to find similar solutions starting with the superpotential
\begin{equation}
W = \mu^N_{Xm} N \overline N_X + \mu^{N}_X N_X \overline N_X + \frac{c_{2,2}}{m_{Pl}} (N \overline N_X)^2 + \frac{c_{n,k}}{m_{Pl}^{2n-3}} (N_X \overline N_X )^{n-k} (N \overline N_X)^k
\end{equation}
where $n\geq 2$ and $k<n$. The third term generates a Majorana mass for the matter right-handed conjugated neutrino, $N$. The full soft-term Lagrangian for this theory is
\begin{align}
-\mc{L}_{soft} = & m_{N}^2 |N|^2 + m_{N_X}^2 |N_X|^2 + m_{\overline N_X}^2 |\overline N_X|^2 - (B\mu^N_{Xm} N \overline N_X +\mbox{ h.c.}) - (B\mu^N_{X} N_X \overline N_X+\mbox{ h.c.}) \nonumber \\
&+ \left(\frac{C_{2,2}}{m_{Pl}}(N \overline N_X)^2+\mbox{ h.c.}\right) +\left(\frac{C_{n,k}}{m_{Pl}^{2n-3}} (N_X \overline N_X )^{n-k} (N \overline N_X)^k + \mbox{ h.c.}\right)
\end{align}
where again $C_{i,j}$ coefficients are $\mc{O}(m_{3/2})$ at the GUT scale.

The F-terms now read
\begin{align}
F_{N} & = \mu^N_{Xm}  \overline N_X  + \frac{2 c_{2,2}}{m_{Pl}} N \overline N_X^2 + \frac{ k c_{n,k}}{m_{Pl}^{2n-3}} N_X^{n-k}  N ^{k-1}\overline N_X^n\\
F_{N_X} & = \mu^N_{X}  \overline N_X  + \frac{ (n-k) c_{n,k}}{m_{Pl}^{2n-3}} N_X^{n-k-1}  N ^{k}\overline N_X^n\\
F_{\overline N_X} &= \mu^N_{Xm} N  + \mu^N_{X} N_X  + \frac{2 c_{2,2}}{m_{Pl}} N^2 \overline N_X + \frac{n c_{n,k}}{m_{Pl}^{2n-3}} N_X^{n-k} N^k \overline N_X^{n-1}
\end{align}
which have a significantly more challenging look than the simplified version presented above. Nonetheless, the same conclusions hold. The above F-terms become more tractable for the $k=0$ and $k=n-1$ cases. In these cases it is possible to get algebraic expressions for the VEVs estimates. For the $k=0$, the F-flatness conditions alone give us
\begin{align}
	N \overline N_X &= - \frac{\mu^N_{Xm} m_{Pl}}{2 c_{2,2}} \\
	N_X \overline N_X &= \left(- \frac{\mu^N_X m_{Pl}^{2n-3}}{n c_{n,o}}\right)^{\frac{1}{n-1}}
\end{align}
while for $k=n-1$, analogous expressions can be obtained
\begin{align}
	| N \overline N_X | &\simeq (  \mu^N_{Xm} m_{Pl}^{2n-3} 	)^{\frac{1}{n-1}} \\
	|N_X \overline N_X| &\simeq ((\mu^N_X)^{3-n} m_{Pl}^{3n-5})^{\frac{1}{n-1}}
\end{align}
where the approximations mean we dropped $\mc{O}(1)$ parameters and took all $\mu$-terms to be of the same order, which is expected.

In both cases, the ratio between the $N_X$ and $N$ VEV is follows the same dependency on $n$
\begin{equation}
\left|\frac{N_X}{N}	\right| \simeq \left(\frac{m_{Pl}}{\mu}\right)^{\frac{n-2}{n-1}}\simeq \begin{cases}
1 & n=2 \\
10^{7.5} & n=3 \\
10^{10} & n = 4
\end{cases}
\end{equation}
where we $\mu$ is an $\mc{O}(\mu^N_{X},\mu^N_{Xm})$ parameter. This result shows that there is a hierarchy between $N_X$ and $N$ VEVs, which is very desirable as $N$ VEVs can generate large B-RPV couplings, c.f. Section \ref{sec:RPV}.

Just like before, we use the D-flat direction
\begin{equation}
\left|\frac{\overline N_X}{N_X}\right|^2=\left| \frac{N}{N_X}\right|^2 + 1,
\end{equation}
which sets the magnitude of the three VEVs. The results for $k=0$ and $k=n-1$ can be immediately estimated algebraically, in contrast to the other cases. The full result of SUSY preserving configurations can be seen in Table \ref{tab:KMSUSYVacua}. It is important to note that for $n=4$, the only viable scenario is for $k=0$, while for $n=3$ the $k=2$ is not viable as there are super-GUT VEVs. In the end we are only interested in the sensible cases, where the VEVs are below the GUT scale and therefore the mechanism is self-consistent.

\begin{table}
	\begin{center}
		{\renewcommand{\arraystretch}{1.5}
			\begin{tabular}{|c|c||c|c|c|}
				\hline\hline
				$n$         & $k$ &     $N$ (GeV)    &    $N_X$  (GeV)   & $\overline N_X$ (GeV)\\ \hline\hline
				\multirow{2}{*}{2} &  0  & $10^{10.5}$ & $10^{10.5}$  &   $10^{10.5}$   \\ \cline{2-5}
				&  1  & $10^{10.5}$ & $10^{10.5}$  &   $10^{10.5}$   \\ \hline\hline
				\multirow{3}{*}{3} &  0  & $10^{6.5}$  & $10^{14.25}$ &  $10^{14.25}$   \\ \cline{2-5}
				&  1  & $10^{10.2}$ & $10^{15.5}$  &   $10^{15.5}$   \\ \cline{2-5}
				&  2  & $10^{10.5}$ &  $10^{18}$   &    $10^{18}$    \\ \cline{2-5}\hline\hline
				\multirow{4}{*}{4} &  0  & $10^{5.5}$  & $10^{15.5}$  &   $10^{15.5}$   \\ \cline{2-5}
				&  1  & $10^{10.1}$ & $10^{16.5}$  &   $10^{16.5}$   \\ \cline{2-5}
				&  2  & $10^{10.3}$ &  $10^{18}$   &    $10^{18}$    \\ \cline{2-5}
				&  3  & $10^{10.5}$ & $10^{20.5}$  &   $10^{20.5}$   \\ \cline{2-5}\hline\hline
			\end{tabular} }
		\end{center}
		\label{tab:KMSUSYVacua}
		\caption{Estimate of the magnitude of the VEVs in SUSY vacua for different implementations of the modified Kolda-Martin mechanism. In all cases the scalar component of the (CP conjugated)  
		right-handed neutrino field $N$ develops a VEV, breaking R-parity, in addition to the 
		$N_X$ and $\overline{N}_X$ VEVs.}
	\end{table}

The SUSY configurations above are expected stabilise the soft-terms Lagrangian just before. The stabilisation conditions are
\begin{align*}
m_{\Phi_1}^2 \Phi_1^*  - B\mu_1  \bar \Phi  + \frac{2 C_{2,2}}{m_{Pl}}\Phi_1 \bar \Phi^2 &=0 \\
m_{\Phi_2}^2 \Phi_2^*  - B\mu_2  \bar \Phi  +\frac{ n C_{n,0}}{m_{Pl}^{2n-3}} \Phi_2^{n-1}  \bar \Phi^n &=0\\
m_{\bar \Phi}^2 \bar \Phi^*  - B\mu_1 \Phi_1 -B\mu_2 \Phi_2 + \frac{2 C_{2,2}}{m_{Pl}}\Phi_1^2 \bar \Phi +\frac{n C_{n,0}}{m_{Pl}^{2n-3}} \Phi_2^{n}  \bar \Phi^{n-1}&=0
\end{align*}
and re-parametrising the dimensionful soft-terms just as before, the above conditions will resemble the F-flatness conditions in form and so they'll be jointly respected taken the parameters of the theory respect relations between them.

As before, the condition that the above extrema are minima is that the potential has a runaway direction around the origin. This is the same to say that, when close to the origin the potential takes the form
\begin{equation}
 V \simeq {\bf N}^* \cdot M_{ N} \cdot {\bf N}
\end{equation}
with ${\bf N}=(N, N_X, \overline N_X^*)$, such that $M_{N}$ at least one negative eigenvalue to account for a run-away behaviour at the trivial extremum. Boundness of the potential in the D-flat direction is achieved by noticing that -- for each field direction -- at least a quadratic term from the non-renormalisable interactions becomes the leading contribution, while keeping a run-away behaviour at the origin.

%================================================================%
\section{Neutrino-neutralino mass matrix}\label{neutrino_section}
%================================================================%

The different breaking scenarios discussed in the previous section rely on different superpotential terms, which are either present or suppressed depending the discrete symmetry of the compactified $G_2$ space. Furthermore, the generic presence of a matter field VEV, $\langle N \rangle$, will generate B-RPV terms, as seen in Section \ref{sec:RPV}. In turn, these provide a new source of neutrino masses which has to be taken into account.

To be more precise we enumerate all the interactions that contribute to neutrino masses. First, we let the matter neutrino to have a Yukawa coupling at tree-level, of the form
\begin{equation}
W_{tree} \supset  y_{\nu} N L H_u  \ .
\end{equation}

Next we have to consider the non-renormalizable terms that employ the KM mechanism for each scenario. Alongside this, we also keep a term that can generate a Majorana mass for the matter right-handed conjugated neutrino, $N$. On top of these, we include a set of non-renormalizable terms involving the Higgses or $L$-type fields, in first order of $1/m_{Pl}$. The non-renormalizable terms that will affect the neutral fermion mass matrix are then
\begin{align}\label{eq:Wnonren}
W_{non.ren.}  \supset &  \frac{c_{2,2}}{m_{Pl}}\left(N N\right)(\overline N_X \overline N_X) + \frac{c_{n,k}}{m^{2n-3}_{Pl}} \left(\nx \nxb \right)^{n-k}\left(\nm \nxb \right)^{k} \nonumber \\
&  + \frac{1}{m_{Pl}}\left(b_1 H_d H_u L \overline{L}_X + b_2 L L \overline L_X \overline L_X + b_3 H_d H_u L_X \overline{L}_X + b_4  L  L_X \overline{L}_X \overline{L}_X  \right. \nonumber\\
& +\left. b_5 L_X L_X \overline L_X \overline L_X + b_6 H_d H_u N \overline{N}_X + b_7 L \overline{L}_X N \overline{N}_X +  b_8 L_X \overline{L}_X N \overline{N}_X \right.\nonumber\\
&+ \left. b_9 H_d H_u N_X \overline{N}_X + b_{10} L \overline L_X N_X \overline N_X +
b_{11} L_X \overline L_X  N_X \overline N_X\right) .
\end{align}

The terms that are disallowed by discrete symmetry are generically re-generated as the moduli acquire VEVs. As such, the following K\"ahler potential terms will have an important contribution for neutrino masses
\begin{align}
K \supset & \frac{s}{m_{Pl}} \overline L_X L_X + \frac{s}{m_{Pl}} \overline L_X L + \frac{s}{m_{Pl}} \overline N_X N_X +\frac{s}{m_{Pl}} \overline N_X N +\frac{s}{m_{Pl}} \overline H_u H_d \nonumber \\
& + \frac{s}{m_{Pl}^2}N_X L_X H_u +  \frac{s}{m_{Pl}^2}N L H_u + \frac{s}{m_{Pl}^2}N_X L H_u + \frac{s}{m_{Pl}^2}N L_X H_u + \frac{s}{m_{Pl}^2}\overline N_{X} \overline L_{X} H_d ,
\end{align}
where $s$ denotes a generic modulus fields that counterbalances the discrete charge. This modulus field needs not to be the same for each coupling. As the moduli acquire VEVs as they are stabilised, the above terms will generate the effective superpotential
\begin{align}
W_{eff} \supset & \mu^L_{XX} \overline{L}_X  L_X +  \mu^L_{Xm} \overline{L}_X  L+\mu^N_{XX} \overline{N}_X N_X+\mu^N_{Xm} \overline{N}_X N+ \mu H_u H_d\nonumber\\
&+ \lambda_{\overline X \overline X} H_d \overline{L}_X \overline{N}_X + \lambda_\nu H_u L N + \lambda_{mX} H_u L N_X + \lambda_{Xm} H_u L_X  N + \lambda_{XX}  H_u L_X N_X
\end{align}
where the parameters can be estimated to lie inside the orders of magnitude
\begin{align}
	\mu \simeq & m_{3/2}\frac{s }{m_{Pl}} \simeq \mathcal{O}(10^{3})\mbox{ GeV} \\
	\lambda \simeq & m_{3/2}\frac{s }{m_{Pl}^2} \simeq \mathcal{O}(10^{-15}) .
\end{align}

Therefore, the total superpotential, which includes all the interactions that contribute to the neutral fermion mass matrix is give by
\begin{align}\label{eq:Wtotal}
W_{total} & \supset W_{tree} + W_{non.ren.}+ W_{eff} .
\end{align}

 In our framework we have VEVs of the $N$-type fields that can be significantly large, depending on which implementation of the KM mechanism we assume. As such, B-RPV couplings, mixing Higgses superfields with $L$-type superfields, appear in the superpotential as
\begin{equation}
\kappa_m {H}_u L + \kappa_X {H}_u L_X +\kappa_{\overline{X}} {H}_d \overline L_{X}
\end{equation}
where the $\kappa$-parameters read
\begin{align}
\kappa_m &\simeq (y_{\nu} + \lambda_{\nu}) \langle N \rangle +  \lambda_{mX} \langle N_X \rangle \label{eq:kappam}% - \frac{0.1 F^{N_X}}{m_{Pl}} - \frac{0.1 F^{N_m}}{m_{Pl}}
\\
\kappa_X &\simeq   \lambda_{Xm} \langle N \rangle + \lambda_{XX} \langle N_X \rangle\label{eq:kappax}% - \frac{0.1 F^{N_X}}{m_{Pl}} - \frac{0.1 F^{N_m}}{m_{Pl}}
 \\
\kappa_{\overline{X}} &\simeq   \lambda_{\overline{X}\overline{X}}\langle \overline N_{X} \rangle\label{eq:kappaxbar}% - \frac{0.1 F^{N_{\overline{X}}}}{m_{Pl}}
\end{align}
where we are dropping the $F$-terms contribution as the solutions for our KM mechanism presented in Section \ref{sec:sym_breaking_section} are aligned in the $D$ and $F$ directions. We also note that we are assuming no tree-level Yukawa couplings involving extra vector-like $N_X$, $\overline N_X$ for the KM scenarios.

Furthermore, the presence of B-RPV induces a sub-EWS VEV on the scalar components of the $\nu$-type fields. In our case, below the EWS, we expect all $\nu$-type scalars to acquire a non-vanishing VEV, generating a mixing between $N$-type fermions and Higgsinos through
\begin{equation}
\epsilon_m H^0_u N + \epsilon_X H^0_u N_X + \epsilon_{\overline{X}} H^0_d N_{\overline{X}}
\end{equation}
where the coefficients read
\begin{align}
\epsilon_m &\simeq (y_{\nu} + \lambda_{\nu} ) \langle \nu \rangle +  \lambda_{mX} \langle \nu_X \rangle \\
\epsilon_X &\simeq \lambda_{Xm} \langle \nu \rangle +  \lambda_{XX} \langle \nu_X \rangle \\
\epsilon_{\overline{X}} &\simeq \lambda_{\overline{X}\overline{X}} \langle \nu_{\overline{X}} \rangle
\end{align}
and, as expected, they have the same generic form as the $\kappa$-parameters since both set of parameters arise from trilinear, Yukawa, couplings in the superpotential.

Finally, as in the MSSM, the presence of VEVs will mix some fermions with gauginos through kinetic terms, namely the Higgsinos with $\tilde B_1$, $\tilde W^0$  due to the Higgses VEVs. In our case we also have $N$-type and $\nu$-type scalar VEVs, which will mix gauginos with matter fermions through kinetic terms. We have, for the $SU(2)$ states,
\begin{equation}
g' \widetilde{B} \langle \widetilde{\nu}_i\rangle  \nu_i,\;\;\; g \widetilde{W}^0 \langle \widetilde{\nu}_i\rangle  \nu_i,\;\;\; g'' \widetilde{B}_X \langle \widetilde{\nu}_i\rangle \nu_i
\end{equation}
while for the $N$-states, which are singlets under the SM gauge group, the mixing with the gaugino of the extra $U(1)_X$ gauge group
\begin{equation}
g'' \widetilde{B}_X \langle \widetilde{N}_i \rangle N_i
\end{equation}
where, in both expressions, we used the shorthand $g' = \sqrt{\frac{5}{3}} g_1$ and $g'' = \frac{1}{2\sqrt{10}}g_X$.

With all the above considerations, we can now construct the $11\times 11$ mass matrix for neutral fermions of our model. We define this matrix in the basis
\begin{equation}
\psi = (\widetilde{B},\widetilde{W}^0,\widetilde{B}_X, \widetilde{H}^0_d,\widetilde{H}^0_u,\nu,\nu_X,\nu_{\overline{X}},N,N_X,N_{\overline{X}}),
\end{equation}
and it has the schematic form
\begin{equation}
\mathbf{M}_{\chi-\nu} = \begin{pmatrix}
\mathbf{M}_{\chi^0}^{5\times5} & \mathbf{M}_{\chi\nu}^{5\times6} \\
(\mathbf{M}_{\chi\nu}^{5\times6})^T & \mathbf{M}_{\nu}^{6\times6}
\end{pmatrix}. \label{mass_matrix}
\end{equation}

The usually called neutralino part of the matrix includes only mass terms involving gauginos and Higgsinos, and its form is very similar to the MSSM, except we have an extended gauge group with one more $U(1)_X$ factor. It reads
\begin{equation}
\mathbf{M}_{\chi^0}^{5\times5} = \begin{pmatrix}
M_1& 0 & 0 & -\frac{1}{\sqrt{2}}g' v_d & \frac{1}{\sqrt{2}}g' v_u\\
0 & M_2 & 0 & \frac{1}{\sqrt{2}}g v_d & -\frac{1}{\sqrt{2}}g v_u\\
0 & 0 & M_X & -2\sqrt{2} g'' v_d & 2\sqrt{2} g'' v_u \\
-\frac{1}{\sqrt{2}}g' v_d & \frac{1}{\sqrt{2}}g v_d & -2\sqrt{2} g'' v_d & 0 & -\mu\\
\frac{1}{\sqrt{2}}g' v_u & -\frac{1}{\sqrt{2}}g v_u & 2\sqrt{2} g'' v_u & -\mu &0
\end{pmatrix}
\end{equation}

The next block is the one involving terms mixing the usual neutralino states with matter states. As such, they include B-RPV masses that mix matter with higgses. The matrix reads
\begin{equation}
\mathbf{M}_{\chi\nu}^{5\times6} = \begin{pmatrix}
-\frac{1}{\sqrt{2}}g' N & -\frac{1}{\sqrt{2}}g' N_X & \frac{1}{\sqrt{2}}g' \overline N_X & 0 & 0 & 0\\
\frac{1}{\sqrt{2}}g N & \frac{1}{\sqrt{2}}g N_X & -\frac{1}{\sqrt{2}}g \overline N_X & 0 & 0 & 0\\
3\sqrt{2}g'' N & 3\sqrt{2}g'' N_X & -3\sqrt{2}g'' \overline N_X & -5\sqrt{2}g'' \nm  & -5\sqrt{2}g'' \nx  & 5\sqrt{2}g'' \nxb \\
0 & 0 & \kappa_{\overline{X}} & 0 & 0 &\epsilon_{\overline{X}}\\
\kappa_m & \kappa_X & 0 & \epsilon_m & \epsilon_X & 0
\end{pmatrix}
\end{equation}
where, in order to de-clutter notation, we are taking the fields names as to represent the VEVs. We notice that the B-RPV couplings $\kappa$ and $\epsilon$ are superpotential terms, while the top three rows is generated by kinetic terms only.

The lower-right $6\times 6$ block is purely from the superpotential, and includes only the masses involving $\nu$-type and/or $N$-type fermions. To obtain the mass, one performs the usual SUSY rule for fermionic masses
\begin{equation}
 \left(\mathbf{M}_{\nu}^{6\times6}\right)_{ij} = -\frac{1}{2}\frac{\partial^2}{\partial \psi_i \partial \psi_j} W_{total}
\end{equation}
where $i$, $j=\{\nu, \nu_X, \overline \nu_X, N, N_X, \overline N_X\}$.

This $6 \times 6$ matrix has three main blocks: the $\nu \nu$ block, $\nu N$ block, and $N N$ block. Schematically they are arranged, in our basis, as
\begin{equation}
\mathbf{M}_{\nu}^{6\times6} = -\frac{1}{2}
\left(
\begin{array}{c|c}
 M_{\nu\nu} & M_{\nu N} \\
  \hline
M_{\nu N}^T &  M_{NN}
\end{array}
\right)
\end{equation}

The actual form of the matrix is obtained using the full superpotential in Eq. \eqref{eq:Wtotal}. Doing so, one gets the following sub-blocks. First we have the $\nu\nu$ block that has mixing between $\overline \nu_X$ and $\nu$, $\nu_X$. In the sub-basis $(\nu, \nu_X, \overline \nu_X)$ this reads
\begin{equation}\label{eq:Mnunu}
M_{\nu\nu}=
	\begin{pmatrix}
		0 & 0 & \frac{b_7 \overline N_X N}{m_{Pl}}+\frac{b_{10} \overline N_X N_X}{m_{Pl}}+\mu^L_{Xm} \\
		  & 0 & \frac{b_8 \overline N_X N}{m_{Pl}}+\frac{b_{11} \overline N_X N_X}{m_{Pl}}+\mu^L_X    \\
		  &   & 0
	\end{pmatrix}
\end{equation}
where we dropped the terms $ \nu^2/m_{Pl}$, $v_{u/d}^2/m_{Pl}$ as they are irrelevant and to de-clutter, and since this block is symmetric we omit the lower left triangular part. But notice that the terms with coefficients $b_7$, $b_8$, $b_{10}$, $b_{11}$ can play an important role as they can generate heavy Dirac masses, depending on the KM mechanism.

Next we have the $\nu N$ block, where one can find the neutrino Dirac masses generated by the Higgses VEV at the EWS. Taking the rows to be along the basis $(\nu, \nu_X, \overline \nu_X)$, while the columns along $(N, N_X, \overline N_X)$, this block reads
\begin{equation}
M_{\nu N}=
\begin{pmatrix}
	v_u y_\nu+\frac{b_7 \overline N_X \overline \nu_X}{m_{Pl}}                   & \frac{b_{10} \overline N_X \overline \nu_X}{m_{Pl}}               & \frac{b_7 N \overline \nu_X}{m_{Pl}}+\frac{b_{10} N_X \overline \nu_X}{m_{Pl}}                                                                                \\
	\frac{b_8 \overline N_X \overline \nu_X}{m_{Pl}}            & \frac{b_{11} \overline N_X \overline \nu_X}{m_{Pl}}               & \frac{b_8 N \overline \nu_X}{m_{Pl}}+\frac{b_{11} N_X \overline \nu_X}{m_{Pl}}                                                                                \\
	\frac{b_7 \overline N_X \nu }{m_{Pl}}+\frac{b_8 \overline N_X \nu_X}{m_{Pl}} & \frac{b_{10} \overline N_X \nu }{m_{Pl}}+\frac{b_{11} \overline N_X \nu_X}{m_{Pl}} & \frac{b_7 N \nu }{m_{Pl}}+\frac{b_{10} N_X \nu }{m_{Pl}}+\frac{b_8 N \nu_X}{m_{Pl}}+\frac{b_{11} N_X
 	\nu_X}{m_{Pl}}
\end{pmatrix}
\end{equation}
where we dropped the sub-leading terms $v_{u/d} \lambda \simeq \mathcal{O}(10^{-12})$ GeV.

Finally we have the $N N$ block, that involves Dirac and Majorana masses generated through the first two terms in Equation \eqref{eq:Wnonren}. Ignoring the terms generated by Higgses and sneutrino VEVs, in the sub-basis $(N, N_X, \overline N_X)$ this block reads
{\tiny
\begin{align}
	&M_{N N} = \\
	& \begin{pmatrix}
		\frac{c_{n,k} (k-1) k  }{ m_{Pl}^{2n-3}} \overline N_X^{n}  N_X^{n-k}   N^{k-2}+\frac{2   c_{2,2}}{m_{Pl}} \overline N_X^2     & \frac{ c_{n,k} k  (n-k)   }{ m_{Pl}^{2n-3}}  \overline N_X^{n} N^{k-1}  N_X^{-k+n-1}   & \frac{c_{n,k} k n}{m_{Pl}^{2n-3}}   \overline N_X^{n-1} N_X^{n-k} N^{k-1}+\mu^N_{Xm}+\frac{4   c_{2,2}}{m_{Pl}}  \overline N_X N \\
		                                             & \frac{ c_{n,k} (-k+n-1) (n-k)} {m_{Pl}^{2n-3}} \overline N_X^{n} N^k  N_X^{n-k-2}      & \frac{c_{n,k}n (n-k)}{m_{Pl}^{2n-3}}  \overline N_X^{n-1} N_X^{-k+n-1} N^k+\mu^N_{XX}                                            \\
		 &  & \frac{c_{n,k}(n-1) n}{m_{Pl}^{2n-3}}  \overline N_X^{n-2} N_X^{n-k} N^k+\frac{2  c_{2,2}}{m_{Pl}}N^2
	\end{pmatrix} \nonumber
\end{align} }
where the orders of magnitude of each entry will largely depend on which KM scenario is being considered. The matrix is symmetric so only the upper diagonal entries are displayed.

%================================================================%

\subsection{The mass matrix hierarchies}\label{sec:Mhierarchies}

	Following the description of the mass matrix above, we will now try to infer the hierarchies between the entries of the matrix. First we notice that, regardless of the case (i.e. the allowed 
	Kolda-Martin operators), the biggest entry in the mass matrix is always in the Gaugino-$N$ mixing block. \footnote{The caveat to this statement is if we allow for an order 1 Neutrino Yukawa, in that case the $\kappa$ entry originated from $ y_\nu \langle N \rangle L H_u $, will have the same order of magnitude. But since the B-RPV coupling above does not involve $\tilde B_X$, $N$, $N_X$, or $\overline{N}_X$, the magnitude of this coupling does not change the following discussion. We will return to B-RPV couplings further below.} This result is understandable as we expect the breaking of the extra $U(1)_X$ to transform a chiral superfield and a massless vector superfield into a single massive vector superfield. The degrees of freedom add up correctly, and would mean that below the $U(1)_X$ breaking scale we can take $\tilde B_X$ and the linear combination of $N$-states that break the $U(1)_X$ to be integrated out jointly. The linear combination that breaks the extra $U(1)_X$ depends on the exact values of the VEVs, but we can highlight some characteristics and how the mass-matrix will look like after this is integrated out.

	In order to single out the correct liner combination that breaks the extra $U(1)_X$, one can perform a rotation in the last three states -- $N$, $N_X$, $\overline N_X$ -- in order to retain only one mixing mass between these states and the $\tilde B_X$. In order to do so, in the limit the mass matrix is real, the rotation is
	\[ U =
	\begin{pmatrix}
	1 & 0      & &                &                              &  \\
	0 & 1      & \cdots&         &                        &                          \\
	& \vdots & \ddots &         &                              &  \\
	&  & &\cos (\theta ) & - \sin (\theta) \cos (\phi ) & \sin
	(\theta ) \sin (\phi )  \\
	&  & &\sin (\theta ) & \cos (\theta ) \cos (\phi )  & -\cos
	(\theta ) \sin (\phi ) \\
	&  & &0              & \sin (\phi )                 & \cos (\phi )
	\end{pmatrix}
	\]
	where the angles are determined by the strength of the mixing mass parameters. For instance, in the $n=2,k =0$ Kolda-Martin mechanism presented before, the VEVs of the scalar components of $N$, $N_X$, $\overline N_X$ are all of same order. In such case, taking $\theta \simeq 3 \pi/4$ and $\phi  \simeq \arctan \sqrt{2}$ will leave only one state mixing with $\tilde B_X$. For the other Kolda-Martin implementations, the $N_X$, $\overline N_X$ VEVs are much larger than $N$ VEV and so we can take $\theta \simeq 0$ with $\phi \simeq 3\pi/4$ to accomplish the same.

	The rotation above affects only the last three columns and rows. Since the matrix is unitary (orthogonal in the case the masses are real), the entries of last three columns of a given row will be mixed with at-most order 1 coefficients, and whilst there might be cancellations there will be no order of magnitude enhancements.
Once the rotation is performed one can then integrate out $\tilde B_X$ jointly with its Dirac partner. This in turn will affect all the remainder of the matrix. For example, the entry $i,j$  will receive a contribution from integrating out a Dirac mass at position $a,b$ of order
	\[
	- \frac{M_{i3}M_{bj}}{M_{3b}}
	\]
	with some order one coefficients from the rotation. In this case we are setting one of the indices to 3 as this is the position of $\tilde B_X$ in our basis. The remaining index, $b$, refers to the position of the linear combination that breaks the extra $U(1)_X$. If, for example, the breaking linear combination that breaks the extra $U(1)_X$ is mostly composed of $N_X$, $\overline N_X$ states, the main contribution to the $\nu$ Majorana mass is given by
	\[
	\frac{b_{10}}{m_{Pl}}\langle \nu \rangle \langle \overline \nu_X \rangle  \ll 10^{-10} \mbox{ GeV}
	\]
	even if we let the respective coupling on, i.e. $b_{10} \simeq \mathcal{O}(1)$.
Therefore, after the above rotation and integrating out , the mass matrix remains schematically the same, but with the absence of $\tilde B_X$ and a linear combination composed of $N$, $N_X$, $\overline N_X$. 

	After integrating out the Dirac fermion originated by the breaking, one can see that the Majorana and Dirac masses -- generated at the $U(1)_X$ breaking scale -- involving only the surviving terms of the $N$, $N_X$, $\overline N_X$ system are the leading entries of the mass matrix. These are present in the bottom-right-most $2 \times 2$ block. These states will then be responsible for a type of see-saw mechanism involving the lighter $SU(2)$ doublet states $\nu$, $\nu_X$, $\overline \nu_X$, with EW scale Dirac mass terms. In order to make sense of this see-saw mechanism, the $\nu$-states need to be protected from too much mixing with the remaining gauginos and higgsinos, such that the lightest mass
eigenstate is dominantly composed of $\nu$. Actually 
the mixing between the $\nu$-type states with gauginos is negligible since it is generated by $\nu$-type VEVs and are therefore sub-EWS. But the mixing with Higgsinos is parametrically dependent on $N$-type VEVs through B-RPV terms, the so called $\kappa$ mass parameters.

	The $\kappa$ parameters defined in Equations \eqref{eq:kappam}, \eqref{eq:kappax} and \eqref{eq:kappax} can have other potentially undesirable consequences
		as they can spoil Higgs physics. Take for example the matter B-RPV interaction, with 
		$\kappa_m$ significantly larger than any other mass involving $H_u$. If were to happens, then
		$L$ and $H_u$ superfields would pair up to produce a heavy vector-like pair. Then $H_u$ would be much heavier than the EWS physics and would spoil Higgs physics, where $H_u$ and $H_d$ are identified as a vector-like pair. In order to preserve viable Higgs physics, we need all $\kappa$-parameters to be much smaller than the remaining masses appearing in the Higgs potential.
	
	Finally, there is risk that $\nu$, $\nu_X$, $\overline \nu_X$ states will mix with each other too much. To see this consider the $3\times 3$ sub-block of the matrix as shown in Eq. \eqref{eq:Mnunu}. If all $b_i$ couplings are suppressed, this matrix will maximally mix $\nu$ and $\nu_X$ through the $\mu$-terms. But it is important to note that while most of the $b_i$ interactions will be generated by Higgses and $\nu$-type VEVs (making them naturally sub-leading even if they are allowed by discrete symmetry) there are two terms that can have important contributions
	\beq
	\frac{b_{10}}{m_{Pl}} N_X \overline N_X \nu \overline \nu_X , \ \frac{b_{11}}{m_{Pl}} N_X \overline N_X \nu_X \overline \nu_X,
	\eeq
	which for the KM cases can generate Dirac masses much greater than $\mu$-terms if the respective $b_i$ coefficients are unsuppressed. This can then provide a natural mechanism to split $\nu$ from $\nu_X$, $\overline \nu_X$, if the coupling $b_{10}$ is forbidden while $b_{11}$ is allowed. In this case, we define
	\begin{equation}\label{eq:mu11}
		\mu_{11} = \frac{b_{11}}{m_{Pl}} N_X \overline N_X
	\end{equation}
	and the leading entries for Eq. \eqref{eq:Mnunu}  will take the form
	\beq
	\begin{pmatrix}
		0 & 0 & \mu^L_{Xm}  \\
		0 & 0 & \mu_{11} \\
		\mu^L_{Xm}  & \mu_{11} & 0
	\end{pmatrix}	,
	\eeq
	which will then lead to $\nu_X$, $\overline \nu_X$ to pair up and decouple from $\nu$.

%===========================================================
	\section{Numerical Results}\label{sec:numerical}
%===========================================================
	
	As the full mass matrix presents an intricate structure of relations and hierarchies between different states, it is ultimately impossible to obtain a simple and revealing analytic expression that describes how one should obtain good neutrino physics. Instead, we perform a numerical scan over space, ensuring that the above constraints are satisfied.
	In so doing, we divided the analysis into different realisations of the Kolda-Martin mechanism, parametrised by different values of $(n,k)$, corresponding to the scenarios in 
	Table~\ref{tab:KMSUSYVacua}. 
	
	In all the cases, we considered a point of the parameter space to be good if the mass of the lightest eigenstate of the mass matrix, identified as a physical neutrino, has a mass in the range
	\beq
	[50, 100] \mbox{ meV},
	\eeq
	and in addition that the corresponding eigenstate is mostly composed of 
	the left-handed doublet component $\nu$ (i.e. the state arising from $(\nu \  e)^T$). 
	In order to do so, we compute the decomposition of the eigenstate in the original basis
	\beq
	| \nu_{light} \rangle = \alpha | \nu \rangle + \dots
	\eeq
	and impose $\alpha$ to be the largest of the coefficients. As discussed in the previous section, the prevalence of $\nu$ as the largest component of $\nu_{light}$ will depend greatly on the parameters of the mass matrix that mix different states, i.e. Dirac masses.
For definiteness, we shall also require that the second lightest mass eigenstate (essentially the lightest
non-neutrino-like neutralino) to be at least $100$ GeV.

	For each example, we only allow the particular desired Kolda-Martin operator while preventing all tree-level Yukawas involving states of the extra vector-like family. Furthermore, unless otherwise stated we assume that all quadratic terms in Eq. \eqref{eq:Wnonren} involving large VEVs are turned off. As expected within the $M$ Theory framework, the disallowed tree-level couplings are regenerated through moduli VEVs, and so the respective coupling strength was set to be of $\mathcal{O}(10^{-15})$. Along the same line, the $\mu$-terms generated by moduli VEVs were set to $\mathcal{O}(1)$ TeV.
	
	Below we will show our findings for the only promising cases, which are $(n,k)=(2,0),\ (2,1),\ (3,0)$. The other $(n,k)$ assignments either returned to little points or no viable correlation to enhance $\alpha$. This happens as for the $(3,1)$, $(4,0)$ cases, since $N_X, \ \overline N_X \simeq 10^{15.5}$ GeV, the B-RPV coupling is generically greater than $1$ GeV. As we will see below, the only viable regions of the parameter space coincide with a naturally suppressed B-RPV parameter.

	\subsection{$\nu$ component of the lightest state}
	
	From the discussion above, we expect the value of $\alpha$ to be correlated with some parameters of the theory. Namely, we expect $\alpha$ to be enhanced  if $b_{11}$ is not suppressed and if the B-RPV coupling $\kappa_m$ is much smaller than any other mass involving Higgsinos.
	Since any disallowed tree-level coupling can be regenerated through moduli VEVs with a $\lambda \simeq 10^{-15}$ suppression, we started our numerical study by looking at the behaviour of $\alpha$ as we let  $b_{11}$ vary in the range
	\begin{equation}
	b_{11} \in [10^{-15},1] ,
	\end{equation}
	which, in conjugation with a non-vanishing $N_X$, $\overline N_X$ VEVs will lead to non-vanishing $\mu_{11}$ as defined in Eq. \eqref{eq:mu11}.
	
	In order to assess the strength of the B-RPV term, $\kappa_m$, allowed in the regions of the parameter space that return good neutrinos, we also registered the value of $\kappa_m$ at each point which returned the mass inside the bounds stated.
	
	\vspace{10mm}
	\underline{\bf $(2,0)$ and $(2,1)$ cases}
	\vspace{5mm}
	
	For these two Kolda-Martin implementation cases, the three $(N,N_X,\overline N_X)$ VEVs are all of order $\mathcal{O}(10^{10.5})$ GeV. As such, we allowed these VEVs to take values around
	\begin{equation}
	N,\ N_X,\ \overline N_X \in [10^{9.5},10^{11.5}]\mbox{ GeV}
	\end{equation}
	to cover the range of expected values. Since with these values the mass matrix is very similar for both $(2,0)$ and $(2,1)$ cases, we present them together.
	
	As a consequence of the values of the VEVs above, the $\mu_{11}$ Dirac mass between $\nu_X$, $\overline \nu_X$ , defined in Eq. \eqref{eq:mu11}, will take values spanning
	\begin{equation}
	\mu_{11} = b_{11} \frac{N_X \overline N_X}{m_{Pl}} = b_{11} [10,10^5]\mbox{ GeV}
	\end{equation}
	which means that, only for non-suppressed $b_{11}$ we expect
	\begin{equation}
	\mu_{11} > \mu^L_{Xm}
	\end{equation}
	as required to split $\nu$ from $\nu_X$, as discussed in Section \ref{sec:Mhierarchies}.
	
	The above considerations indicate us that the mechanism to split $\nu$ from $\nu_X$ will only work for large values of $b_{11}$. This can be seen in Figures \ref{fig:(2,0)-Scatter-(alpha,b11)} and \ref{fig:(2,1)-Scatter-(alpha,b11)}, where a slight agglomeration of points around $(\alpha,b_{11})\simeq (1,1)$ can be identified.
	
	\begin{figure}[h]
		\centering
		\subfigure[$(2,0)$ case\label{fig:(2,0)-Scatter-(alpha,b11)}]{\includegraphics[width=0.45\linewidth]{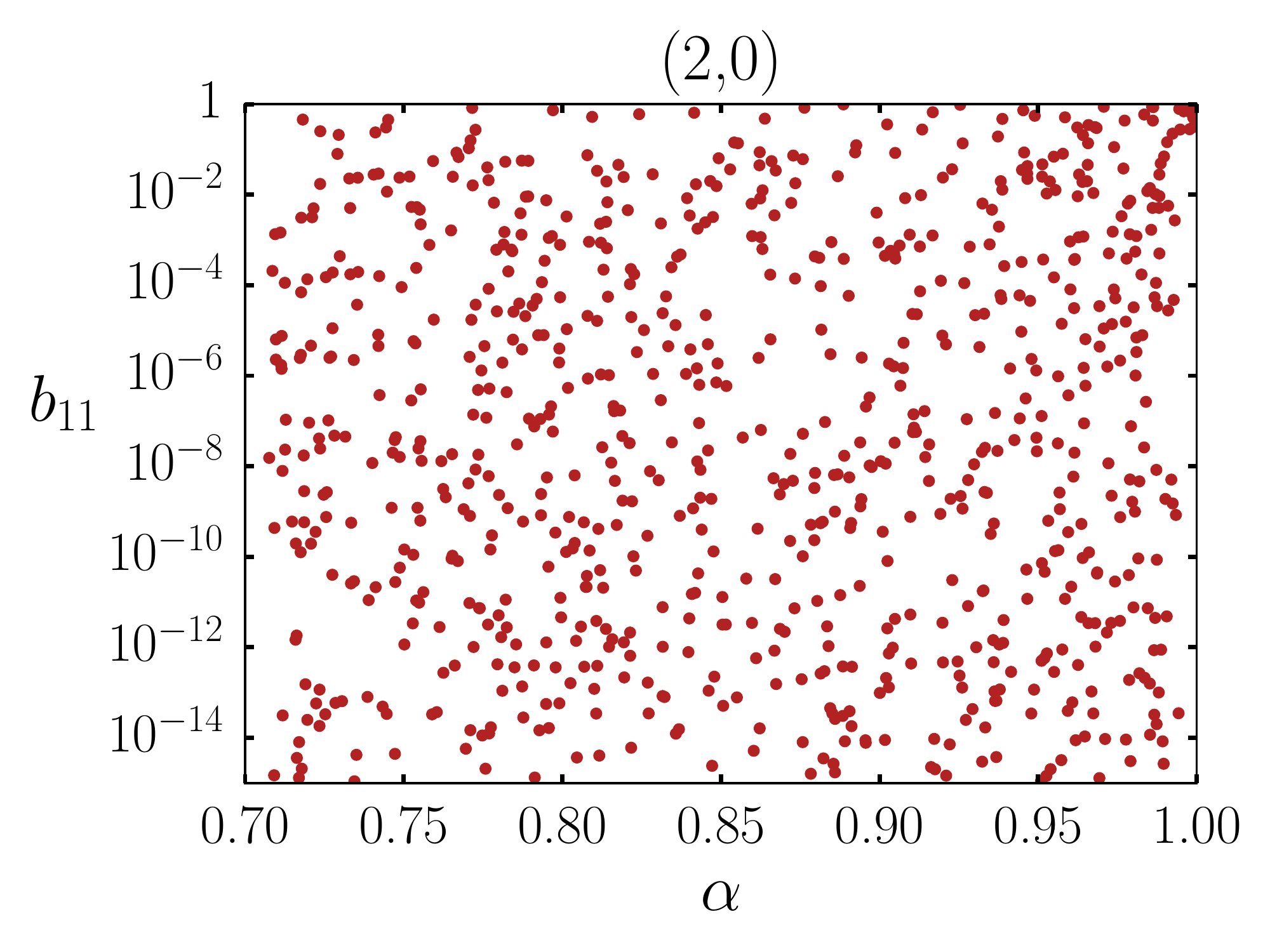}}
		\subfigure[$(2,0)$ case\label{fig:(2,1)-Scatter-(alpha,b11)}]{\includegraphics[width=0.45\linewidth]{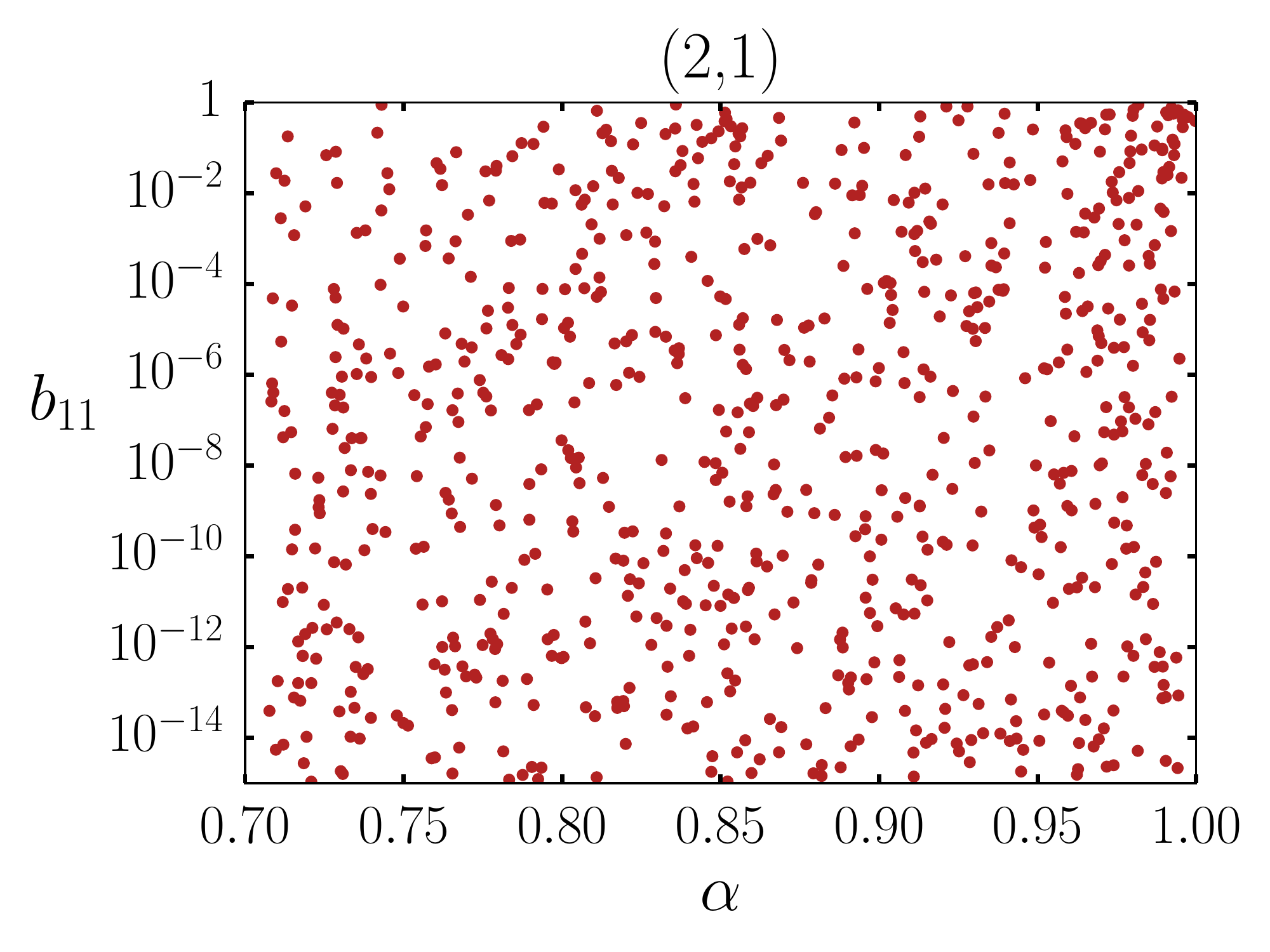}}		
		\caption{Scatter plots showing the amplitude $\alpha$ of the left-handed doublet state
		$\nu$ in the lightest mass eigenstate
		$\nu_{light}$ as $b_{11}$ varies for the $(2,0)$ and $(2,1)$ cases.
		The points are fairly evenly distributed with a slight clustering near the desired value of $\alpha \approx 1$ for $b_{11}\approx 1$.}
	\end{figure}
		
	 On the other hand, we find that the $\kappa_m$ parameter is mostly bounded to be smaller than $1$ GeV, as is shown in Figures \ref{fig:(2,0)-Scatter-(alpha,kappam)} and \ref{fig:(2,1)-Scatter-(alpha,kappam)}. Although such small values of $\kappa_m$ are welcome, the fact that there is no clear preference for $\kappa_m \gtrsim 10^{-2}$ GeV suggests this class of models is challenged by BBN constraints, c.f. Eq. \eqref{eq:kLSPBBN}.
	 
	\begin{figure}[h]
		\centering
		\subfigure[$(2,0)$ case\label{fig:(2,0)-Scatter-(alpha,kappam)}]{\includegraphics[width=0.45\linewidth]{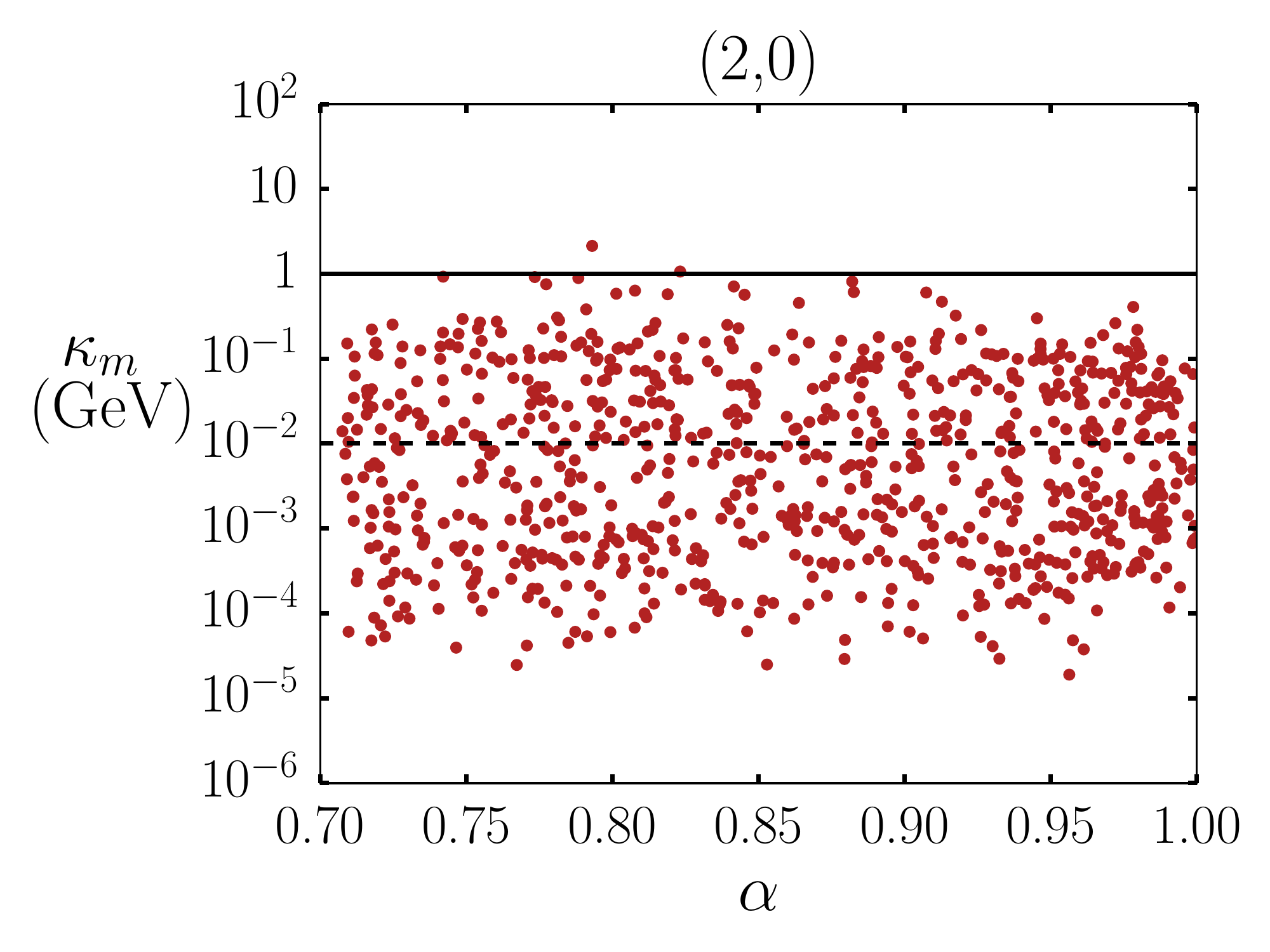}}
		\subfigure[$(2,1)$ case\label{fig:(2,1)-Scatter-(alpha,kappam)}]{\includegraphics[width=0.45\linewidth]{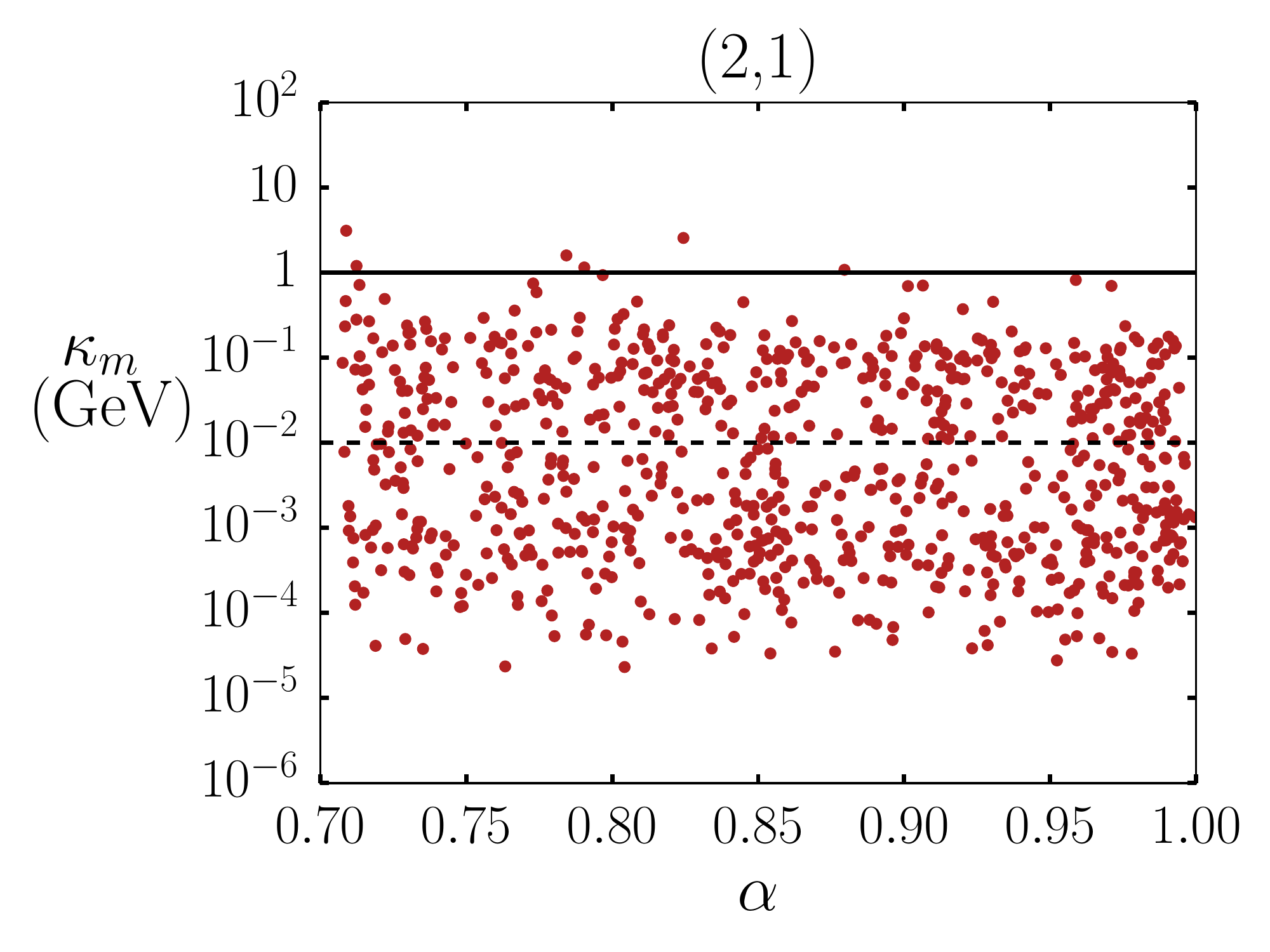}}
		\caption{Scatter plots showing the amplitude $\alpha$ of the left-handed doublet state
		$\nu$ in the lightest mass eigenstate
		$\nu_{light}$ as $\kappa_m$ varies for the $(2,0)$ and $(2,1)$ cases.
		The points are fairly evenly distributed with a slight clustering near the desired value of $\alpha \approx 1$. The horizontal dashed line represents the bound on the LSP lifetime, c.f. Eq. \eqref{eq:kLSPBBN}.}
	\end{figure}

	\vspace{10mm}
	\underline{\bf $(3,0)$ case}
	\vspace{5mm}

	For the $(3,0)$ Kolda-Martin realisation, we found much promising results. Since the $N_X$, $\overline N_X$ VEVs are expected to be around $\mathcal{O}(10^{14.25})$ GeV, if we allow them to be in the range
	\begin{equation}
	N_X,\ \overline N_X \in [10^{13.25},10^{15.25}] \mbox{ GeV}
	\end{equation}
	we find
	\begin{equation}
	\mu_{11} \in b_{11} [10^{8.25},10^{12.25}]\mbox{ GeV}
	\end{equation}
	which implies that it is natural to achieve
	\begin{equation}
	\mu_{11} \gg \mu^L_{Xm}
	\end{equation}
	and consequently $\nu$ will decouple easily from the other $\nu$-type states.
	
	The above expectations are confirmed by the numerical results, and the lightest state will be mostly composed of $\nu$ even for values of $b_{11}$ below $\mathcal{O}(1)$. This behaviour can be seen in Figure \ref{fig:(3,0)-Scatter-(alpha,b11)}.
	
	\begin{figure}[h]
		\centering
		\subfigure[Scatter of $(\alpha,b_{11})$ plane\label{fig:(3,0)-Scatter-(alpha,b11)}]{\includegraphics[width=0.45\linewidth]{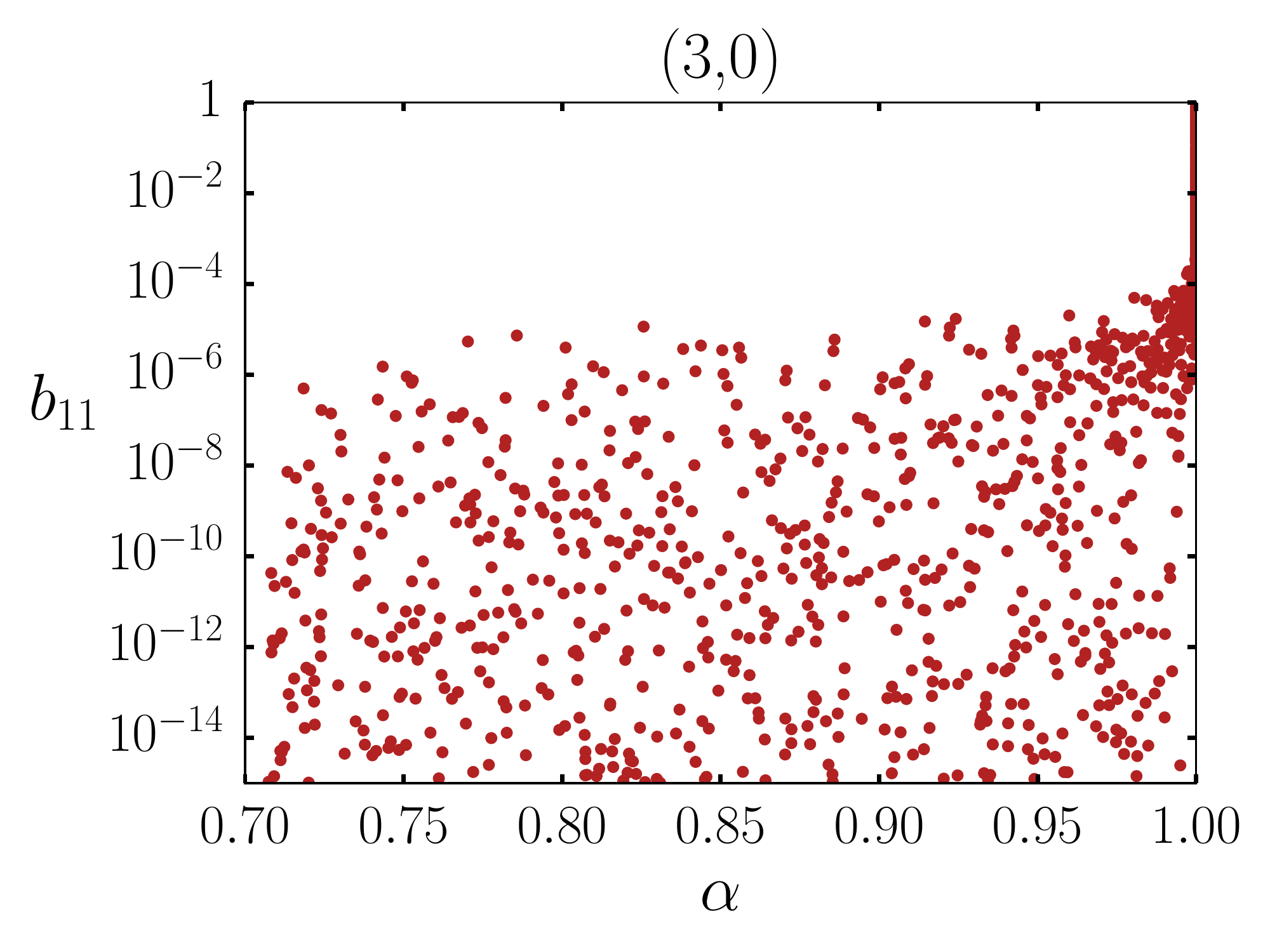}}
		\subfigure[Scatter of $(\alpha,\kappa_m)$ plane\label{fig:(3,0)-Scatter-(alpha,kappam)}]{\includegraphics[width=0.45\linewidth]{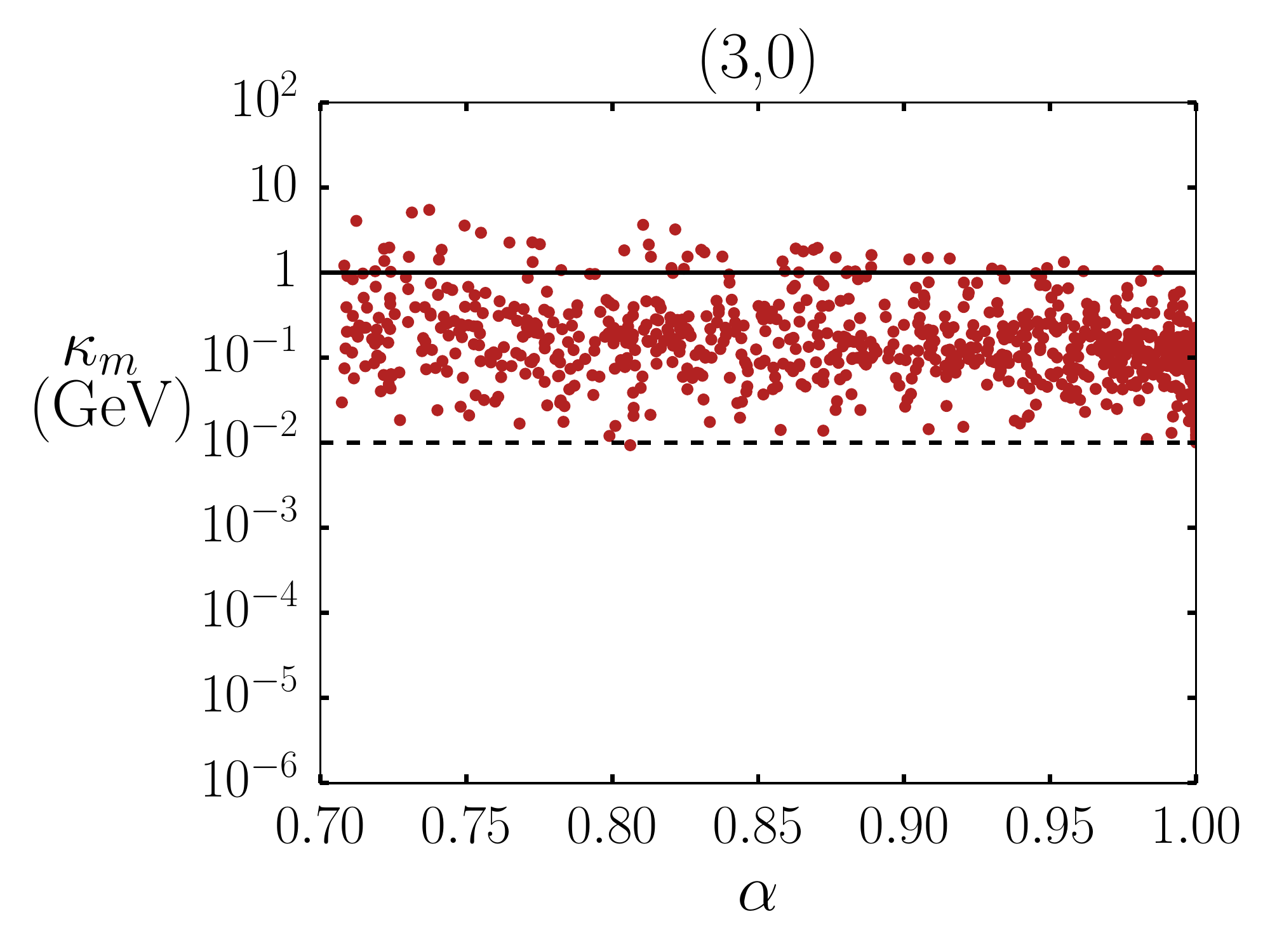}}
		\caption{Scatter plots showing the amplitude $\alpha$ of the left-handed doublet state
		$\nu$ in the lightest mass eigenstate
		$\nu_{light}$ as $\kappa_m$ varies for the $(3,0)$ case.
		The points are fairly evenly distributed except for a significant clustering near the desired value of $\alpha \approx 1$ for larger values of $b_{11}$. The horizontal dashed line represents the bound on the LSP lifetime, c.f. Eq. \eqref{eq:kLSPBBN}. The right panel shows that nearly all the points
		satisfy $\kappa_m \gtrsim 10^{-2}$ GeV.}
	\end{figure}
	
	Interestingly, in the $(\alpha,\kappa_m)$ plane, shown in Figure \ref{fig:(3,0)-Scatter-(alpha,kappam)} we can see again that the mass matrix prefers $\kappa_m < 1$ GeV in order to reproduce a mostly-$\nu$ lightest state. This is a nice result which ensures that whenever we have good physical neutrinos, we also find sufficiently suppressed B-RPV.
	Furthermore, all the good points also suggest $\kappa_m \gtrsim 10^{-2}$ GeV, 
	satisfying the requirement for successful BBN physics, c.f. Eq. \eqref{eq:kLSPBBN}.
	
	\subsection{Matter Neutrino Yukawas and B-RPV couplings}

	From the above analysis we learned that for the $(2,0)$, $(2,1)$ and $(3,0)$ cases we expect a non-suppressed $b_{11}$ to enhance the component of $\nu$ in the lightest state. As such, we will now consider this coupling to be of order 1 and re-run the analysis for these cases, with the goal being to assess what typical values $\kappa_m$ and $y_\nu$ should take for a successful implementation of the proposed Kolda-Martin mechanism.

	\vspace{10mm}
	\underline{\bf $(2,0)$ and $(2,1)$ cases}
	\vspace{5mm}	
	
	In Figures \ref{fig:(2,0)-Histogram-Ynu} and \ref{fig:(2,1)-Histogram-Ynu} we see that the preferred points are those with $y_\nu  \lesssim 10^{-10}$. This suggests that for theses cases, the see-saw mechanism does not take a great role in explaining the light neutrino masses.
	
	\begin{figure}[h]
		\centering
		\subfigure[$(2,0)$ case\label{fig:(2,0)-Histogram-Ynu}]{\includegraphics[width=0.45\linewidth]{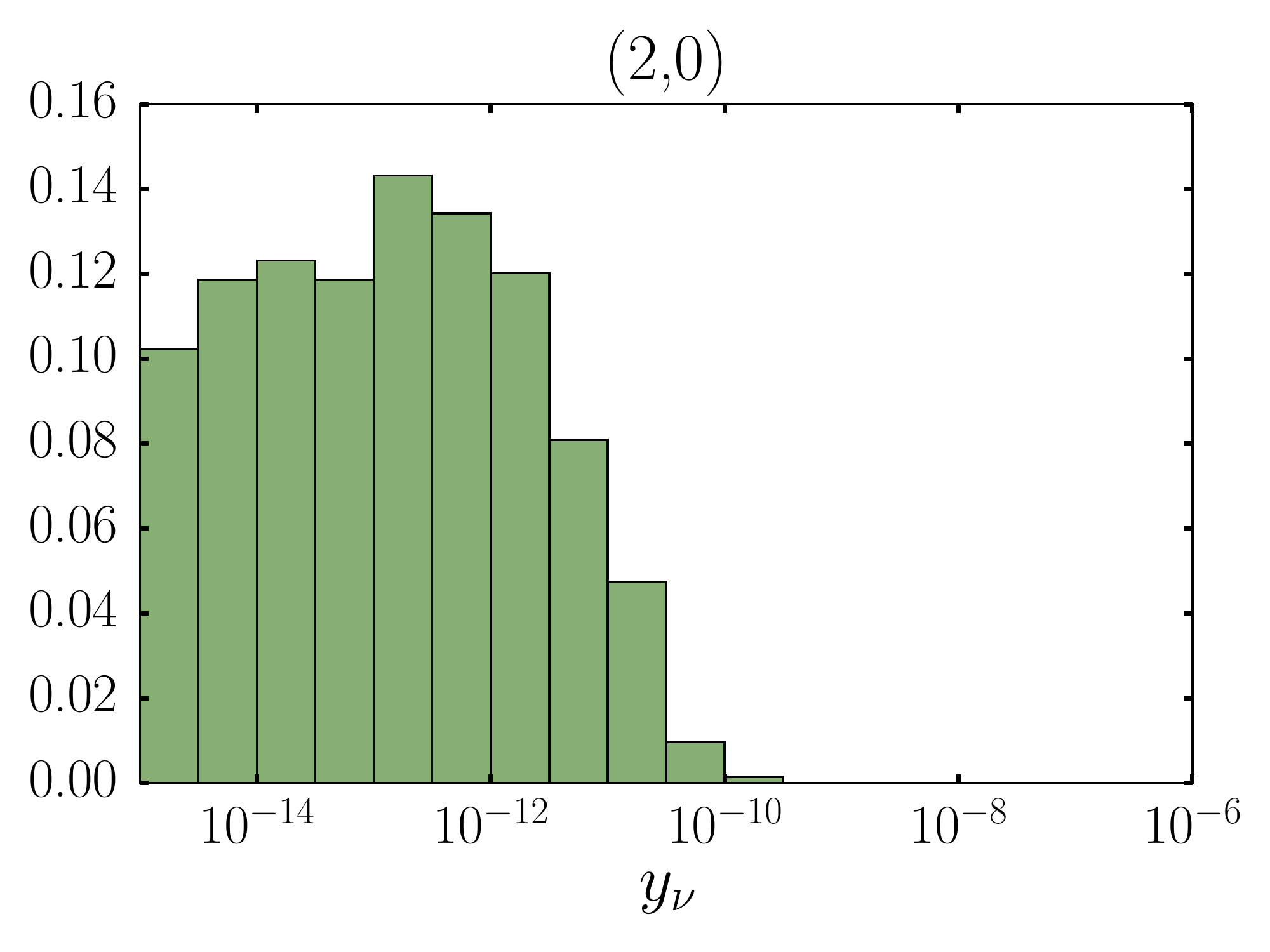}}
		\subfigure[$(2,1)$ case\label{fig:(2,1)-Histogram-Ynu}]{\includegraphics[width=0.45\linewidth]{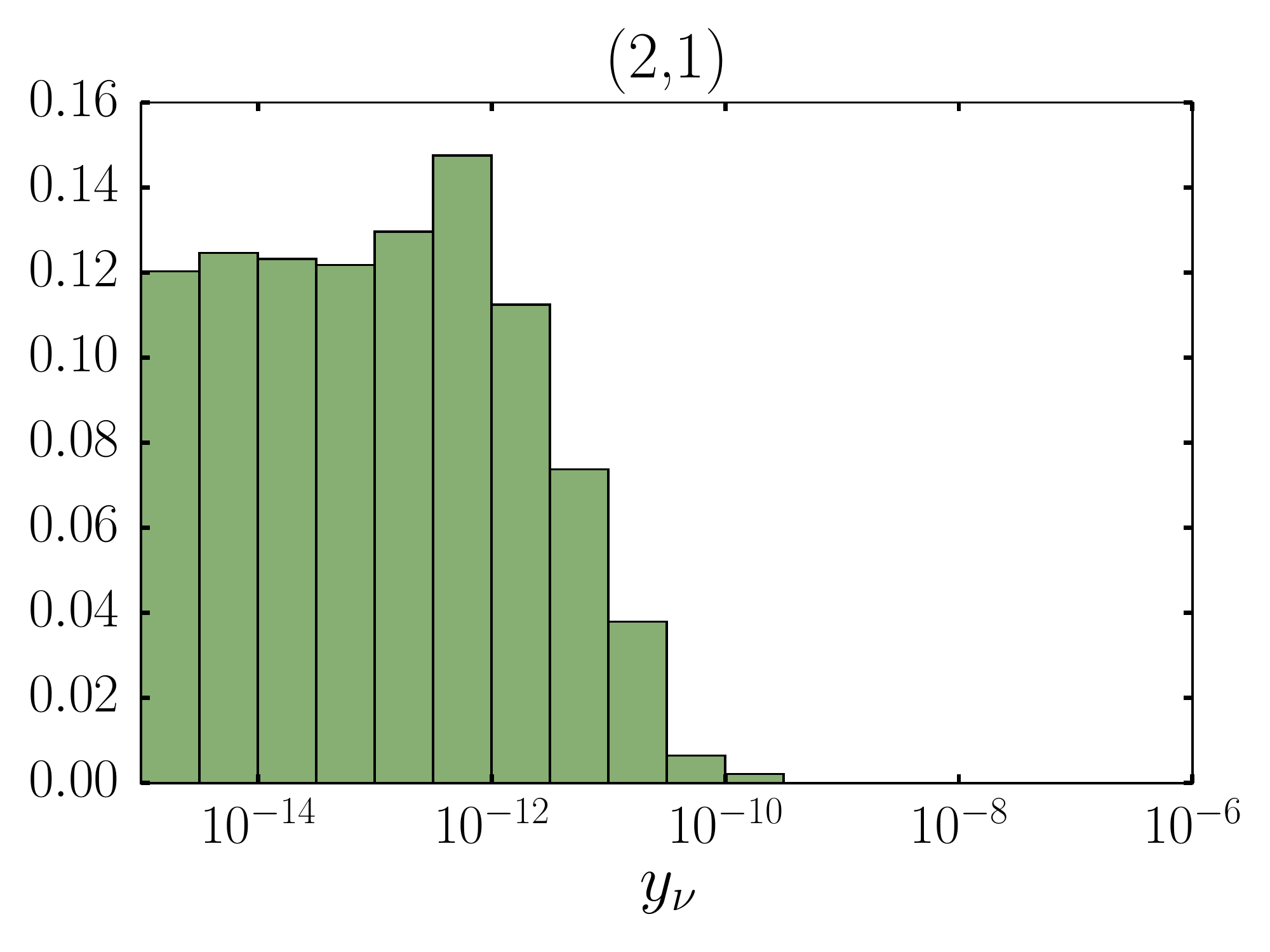}}
		\caption{Histograms for the values of $y_\nu$ for the $(2,0)$ and $(2,1)$ cases with unsuppressed $b_{11}$}
	\end{figure}
	
	In Figures \ref{fig:(2,0)-Histogram-Kappam} and \ref{fig:(2,1)-Histogram-Kappam} we see that for these cases, the B-RPV parameter $\kappa_m$ is naturally very small. This result is easy to understand, considering the main contribution to $\kappa_m$ to be
	\[
	\kappa_m \simeq y_\nu v_m,
	\]
	and given the range of values that we are allowing the VEVs to take, $\kappa_m$ is expected to be small. Unfortunately, all points returning good neutrino physics also return $\kappa_m > 10^{-2}$ GeV, which means that these classes of models spoil BBN, c.f. \eqref{eq:kLSPBBN}.
	Although not shown here one can also find that $\kappa_X$, $\kappa_{\overline X}$ parameters, which mix $L_X$, $\overline L_X$ with $H_u$, $H_d$ respectively, are also constrained to be smaller than 1 $GeV$.
		
	\begin{figure}[h]
		\centering
		\subfigure[Histogram for values of $\kappa_m$\label{fig:(2,0)-Histogram-Kappam}]{\includegraphics[width=0.45\linewidth]{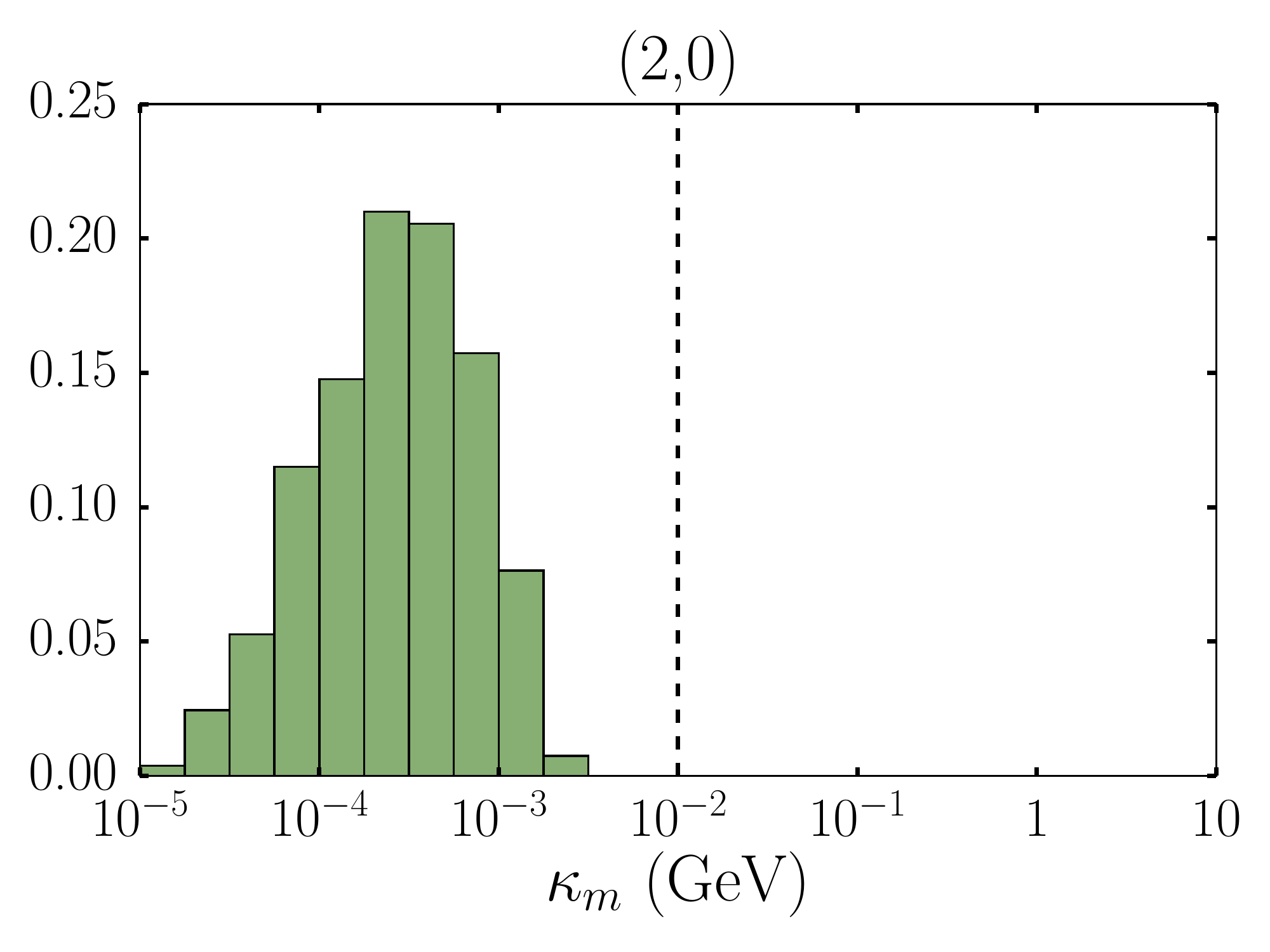}}
		\subfigure[Histogram for values of $\kappa_m$\label{fig:(2,1)-Histogram-Kappam}]{\includegraphics[width=0.45\linewidth]{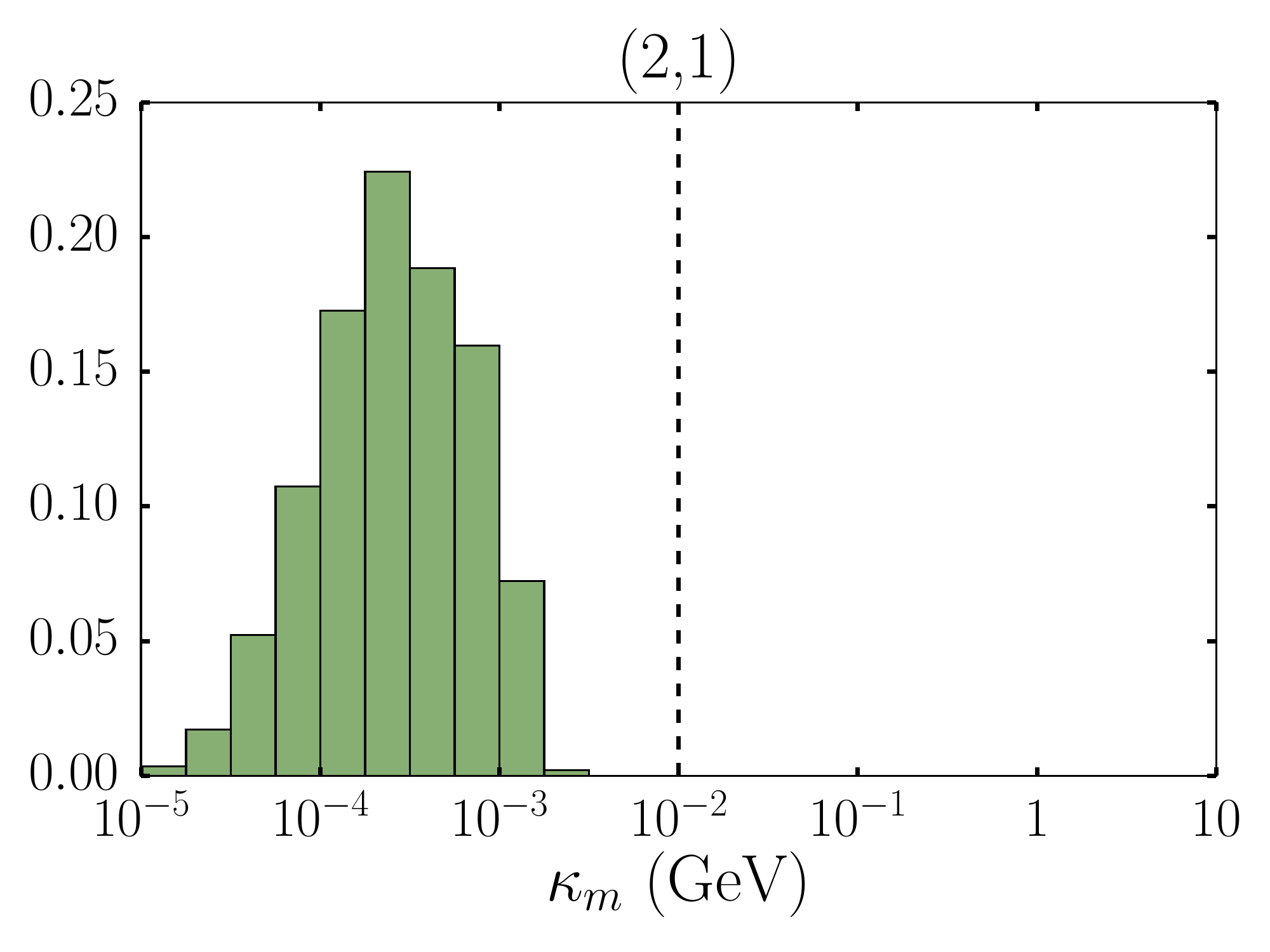}}
		\caption{Histograms for the values of $y_\nu$ and $\kappa_m$  for the $(2,1)$ case with unsuppressed $b_{11}$. The vertical dashed line represents the bound on the LSP lifetime, c.f. Eq. \eqref{eq:kLSPBBN}.}
	\end{figure}

	\vspace{10mm}
	\underline{\bf $(3,0)$ case}
	\vspace{5mm}
	
	For this realisation of the Kolda-Martin mechanism, the results are slightly different but in line with our expectations. In Figure \ref{fig:(3,0)-Histogram-Ynu} we can see that the matter Yukawa coupling is allowed to take values larger than in the previous case. This indicates that the see-saw mechanism is having an effect on reducing the contribution of the matter neutrino Dirac mass to the lightest eigenstate.
	
	\begin{figure}[h]
		\centering
		\subfigure[Histogram for values of $y_\nu$\label{fig:(3,0)-Histogram-Ynu}]{\includegraphics[width=0.45\linewidth]{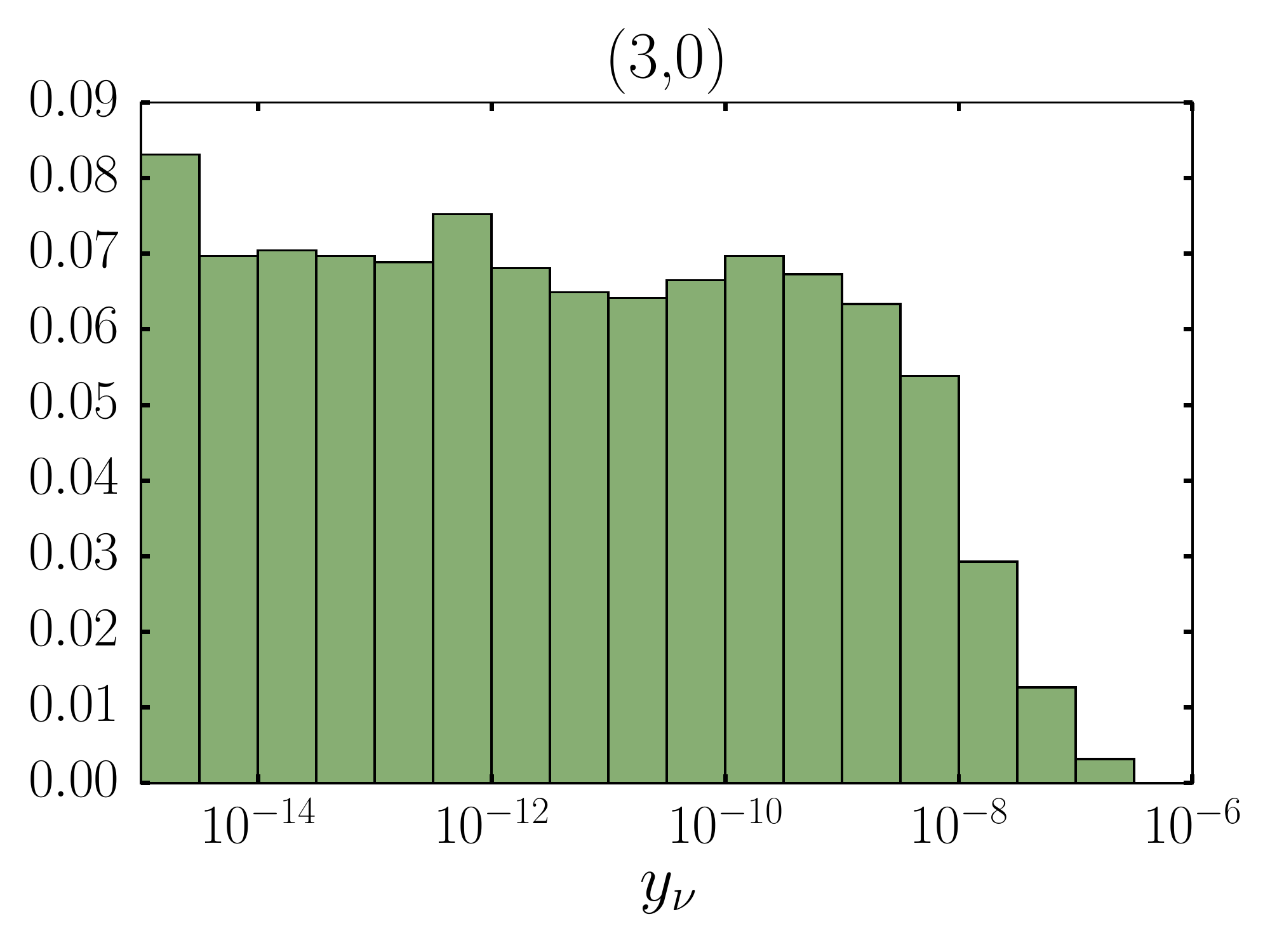}}
		\subfigure[Histogram for values of $\kappa_m$. The vertical dashed line represents the bound on the LSP lifetime, c.f. Eq. \eqref{eq:kLSPBBN}.\label{fig:(3,0)-Histogram-Kappam}]{\includegraphics[width=0.45\linewidth]{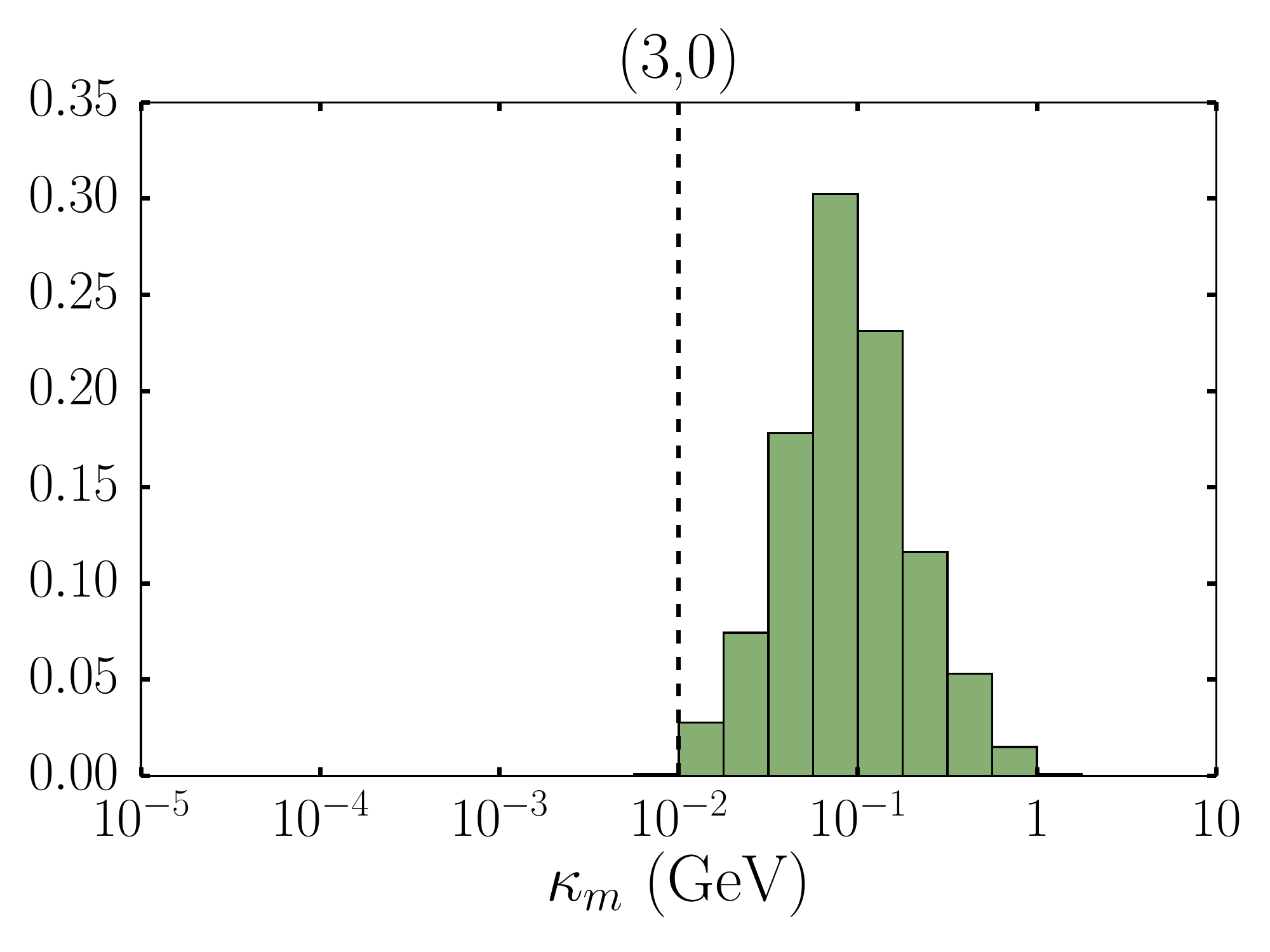}.pdf}
		\caption{Histograms for the values of $y_\nu$ and $\kappa_m$  for the $(3,0)$ case with unsuppressed $b_{11}$}
	\end{figure}

	In Figure \ref{fig:(3,0)-Histogram-Kappam} we see that $\kappa_m$ is bound to be smaller than $1$ GeV. The fact that $\kappa_m$ takes larger values for $(3,0)$ case than for the $n=2$ cases is easily understandable. The main contributions to $\kappa_m$ are
	\begin{equation}
	\kappa_m \simeq y_\nu N + \lambda N_X
	\end{equation}
	where the VEVs are expected as in Table \ref{tab:KMSUSYVacua}. These contributions are in general greater than those in $n=2$ cases, but they are still bounded to be smaller than $1$ GeV. This is fortunate, as $\kappa_m \gtrsim 10^{-2}$ GeV and hence this class of models retain the successful predictions of BBN, c.f. \eqref{eq:kLSPBBN}.
As before, although not shown here also finds that $\kappa_X$, $\kappa_{\overline X}$ parameter are also constrained to be smaller than 1 $GeV$.

%================================================================%
\section{Conclusions and Discussion} \label{sec:conclusion}

In this paper we have studied the origin of neutrino mass from $SO(10)$ SUSY GUTs arising from
$M$ Theory compactified on a $G_2$-manifold. We have seen that this problem is linked to 
the problem of $U(1)_X$ gauge symmetry breaking, which appears 
in the $SU(5)\times U(1)_X$ subgroup of $SO(10)$, and remains unbroken by the Abelian Wilson
line breaking mechanism. In order to break the $U(1)_X$ gauge symmetry, we considered a 
(generalised) Kolda-Martin mechanism.
Our results show that it is possible to break the $U(1)_X$ gauge symmetry
without further SUSY breaking while achieving high-scale VEVs that play a crucial role in 
achieving the desired value of neutrino mass.

The subsequent induced R-parity violation provides an additional source
of neutrino mass, in addition to that arising from the seesaw mechanism from non-renormalisable terms. The resulting $11\times 11$ neutrino mass 
matrix was analysed for one neutrino family and it was shown how a
phenomenologically acceptable neutrino mass can emerge. This happens easily for the $(n,k)=(3,0)$ case of the Kolda-Martin mechanism we developed. For this class of models, not only is the neutrino masses phenomenologically viable, but also the physical light neutrino eigenstate is almost entirely composed of the left-handed (weakly charged) state $\nu$ in the same doublet as the electron $(\nu, e)$, as desired.
Furthermore, our analysis showed that the B-RPV parameters, which play an important role in neutrino masses and low-energy dynamics, are in the required range, being smaller than $1$ GeV. Finally, we notice that contrary to the $n=2$ cases, the $n=3$ type of Kolda-Martin mechanism immediately preserves the successful predictions of BBN by allowing the LSP to decay quickly in early universe.

In conclusion, we have shown that $SO(10)$ SUSY GUTs from $M$ Theory on $G_2$ manifolds 
provides a phenomenologically viable framework, in which the rank can be broken in the effective theory
below the compactification scale, leading to acceptable values of neutrino mass,
arising from a combination of the seesaw mechanism
and induced R-parity breaking contributions. 
In principle the mechanism presented here could be extended to three neutrino families and eventually 
could be incorporated into a complete theory of flavour, based on $M$ Theory $SO(10)$,
however such questions are beyond the scope of the present paper.

%%%%%%%%%%%%%%%%%%%%%%%%%%%%%%%%
\section*{Acknowledgements}
The work of BSA is supported by UK STFC via the research grant ST/J002798/1.
SFK acknowledges support from the STFC Consolidated grant ST/L000296/1 and the
European Union Horizon 2020 research and innovation programme under the Marie 
Sklodowska-Curie grant agreements InvisiblesPlus RISE No. 690575 and 
Elusives ITN No. 674896. MCR acknowledges support from the FCT under
the grant SFRH/BD/84234/2012. CP is supported by the KCL NMS graduate school and ICTP Trieste.
The work of KB is supported by a KCL GTA studentship.
%================================================================%
\bibliographystyle{JHEP}
\bibliography{2016Jul}
\end{document}